\documentclass[preprint]{aastex}

\slugcomment{To appear in ApJ}

\shorttitle{Modeling Nova Cygni 1992}
\shortauthors{Short, {\it et al.}}

\begin{document}


\title{Non-LTE Modeling of Nova Cygni 1992}


\author{C. Ian Short}
\affil{Department of Physics, Florida Atlantic University,
    Boca Raton, FL 33431-0991}

\and

\author{Peter H. Hauschildt}
\affil{Department of Physics and Astronomy and Center for
Simulational Physics, University of Georgia,
    Athens, GA 30602-2451}

\and

\author{S. Starrfield}
\affil{Department of Physics and Astronomy, Arizona State University, P.O. Box 871504,
Tempe, AZ 85287-1504 } 

\and

\author{E. Baron}
\affil{Dept. of Physics and Astronomy, University of Oklahoma,
    440 W. Brooks, Rm 131, Norman, OK 73091-0225}



\begin{abstract}

   We present a grid of nova models that have an extremely large number
of species treated in NLTE, and apply it to the analysis of an extensive time
series of ultraviolet spectroscopic data for Nova Cygni 1992.  We use ultraviolet 
colors to derive the time development of
the effective temperature of the expanding atmosphere during the fireball phase
and the first ten days of the optically thick wind phase.  We find that the
nova has a pure optically thick wind spectrum until about 10 days after the explosion.
During this interval, we find that synthetic
spectra based on our derived temperature sequence agree very well with the observed spectra.
We find that a sequence
of hydrogen deficient models provides an equally good fit providing the
model effective temperature is shifted upwards by $\sim 1000$ K.  We find that 
high resolution UV spectra of the optically thick wind phase are fit moderately
well by the models.   We find that a high 
resolution spectrum of the {\it fireball} phase is better fit by a model with a steep density
gradient, similar to that of a supernova, than by a nova model. 
    
\end{abstract}


\keywords{novae, cataclysmic variables --- stars: atmospheres --- stars: individual (Nova Cygni 1992, V1974 Cygni) --- ultraviolet: stars}


\section{Introduction}

    Energetic stellar explosions such as novae and supernovae provide
us with the opportunity to study the physics of rapidly expanding gas shells.  Due to
the rapid response of observers and the density of observational coverage in time,   
the classical \ion{ONeMg}{0} nova Cygni 92 (V1974 Cygni) \citep{collins} is the best observed nova in 
history.  The existence
of a frequent, well spaced set of low and high resolution International Ultraviolet Explorer (IUE)
spectra allow
us to model the time development of the expanding gas shell in a spectral region
where the object emits most of its flux during the fireball and optically thick wind 
phases of the explosion.  The gradual emergence of flux in this region, caused by the lifting 
of heavy line
blanketing (``the iron curtain'' (\citet{ncyg1}, \citet{ncyg2}),
determines the shape of the early UV light curve, and is a particularly dramatic
example of how these explosions serve as laboratories for the study of 
a radiating plasma under changing conditions.
 
\paragraph{}
  
   Among stellar atmospheric modeling problems, that of novae is especially complicated due
to the steepness of the $T$ and $\rho$ gradients, the large depth of translucency, the sphericity, 
and the significance of first order special relativistic effects in the transfer of radiation.
The first two of these conditions cause the line forming region (LFR) to span a wide range
of physical conditions so that many overlapping line and continuum transitions of many ionization 
stages of many elements are simultaneously present in the emergent spectrum.  The second condition
ensures that many species in the LFR will deviate significantly from Local Thermodynamic 
Equilibrium (LTE) \citep{short1}.  As a result, the accurate modeling of data such as 
that obtained for Cygni 92 requires a very general treatment that incorporates non-LTE (NLTE) effects 
as accurately and completely as possible.  The multi-purpose atmospheric modeling code {\tt PHOENIX}
\citep{hausphx} was developed to model, among other things, nova and supernova explosions.
{\tt PHOENIX} makes use of a fast and accurate Operator Splitting/Accelerated Lambda Iteration
(OS/ALI) scheme to solve self-consistently the first order Special Relativistic radiative
transfer equation and the NLTE rate equations for many species and overlapping transitions
\citep{hausphx}.  Cygni 92 was first modeled with {\tt PHOENIX} by \citet{ncyg2}.  However,
recently \citet{short1} have greatly
increased the number of species and ionization stages treated in 
detailed NLTE by {\tt PHOENIX}.  The lowest six ionization stages of the 19 most
important elements are now treated in NLTE.  The purpose of this study is to calculate 
a set of improved nova models and synthetic spectra with the expanded version of {\tt PHOENIX}, 
to compare them to the IUE time series spectra of Cygni 92, and to extract stellar 
parameters at each observed time. 
 
\section{Observations}\label{obs}

\begin{figure}
\plotone{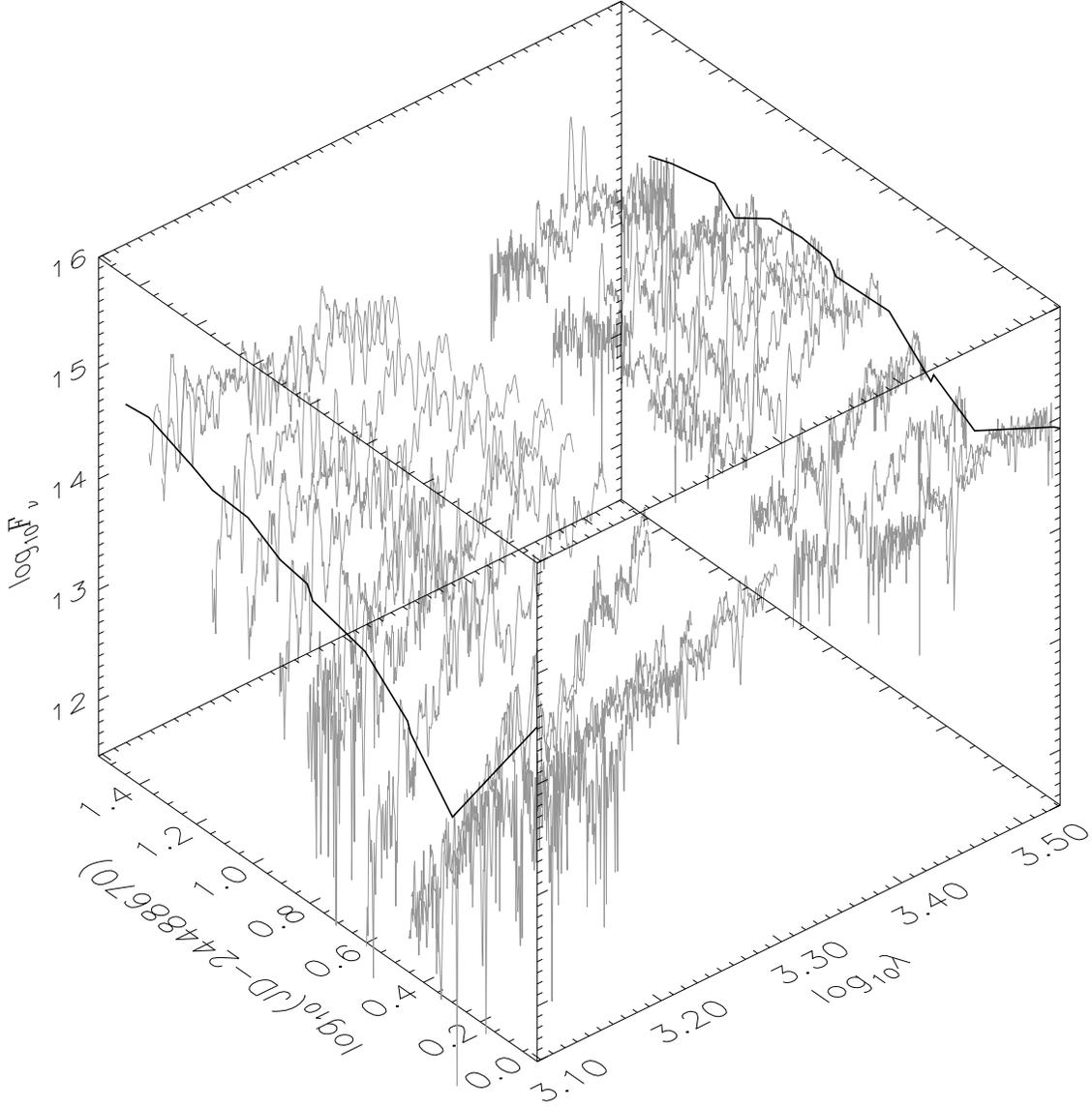}
\figcaption[f1.eps]
{Time series of IUE SWP and LWP spectra at the fourteen phases 
of approximate simultaneity of the bands.  The gap along the $\lambda$ axis
occurs because the reduced SWP and LWP spectra are disjoint in wavelength.  The mean wavelength 
integrated flux, $\bar{F}_\lambda$, for the SWP and LWP bands as a function of time
(light curve) is plotted on the short wavelength and long wavelength 
walls of the plot cube, respectively.  
\label{surf_lo} }
\end{figure}

  Table 1 is a log of the observations that we have extracted from the IUE archive for
both the low and high resolution spectrographs and for both the long and short wavelength cameras 
(LWP and SWP).  Superscripts in Table 1 indicate approximately coincident pairs
of SWP and LWP spectra.  The spectra span a range of 144.75 days, from the initial
fireball stage of the explosion, through the optically thick wind phase, to the nebular 
stage.   Fig. \ref{surf_lo} shows the fourteen pairs of approximately coincident SWP and LWP
spectra as a function of time.  We have discarded the portion of the SWP spectrum that lies below 
$\lambda 1250$ \AA~ due to the noisiness of the signal and contamination by geocoronal Lyman $\alpha$, and the
portion of the LWP spectrum that lies below $\lambda 2400$ due to severe degradation of the signal.  The gap in the
$\lambda$ dimension arises because the reduced SWP and LWP spectra are disjoint in $\lambda$.
The spectra have
been recalibrated in wavelength and flux using the procedure of \citet{massa}, and
de-reddened with a value of $E(B-V)=0.26$ \citep{chochol} using the galactic extinction curve of
\citet{fitz}.

\paragraph{}

   We have calculated the light curves, $\bar{F}_{\rm SWP}(t)$ and $\bar{F}_{\rm LWP}(t)$, for the SWP and LWP 
bands by calculating the wavelength integrated 
mean flux, $\bar{F}_\lambda$, for each flux-corrected, de-reddened spectrum.  The light curves are shown 
projected onto the sides of the data cube in Fig. \ref{surf_lo}.
    From the shape of the light curves and from the 
global shape of the spectra, we can see that the first two spectra were taken during a 
stage of rapid decline in UV flux and rapid softening of the UV spectral energy 
distribution, which is characteristic of the fireball stage of a nova.  Later, 
the UV flux gradually increases and then becomes approximately constant and the spectral distribution gradually 
hardens as the nova progresses through the optically thick wind phase.  Finally, the
latest LWP spectrum in Fig. \ref{surf_lo} shows prominent \ion{Mg}{2} $hk$ emission,
which indicates the onset of the optically thin nebular phase of the nova.  

\paragraph{}

\begin{figure}
\plotone{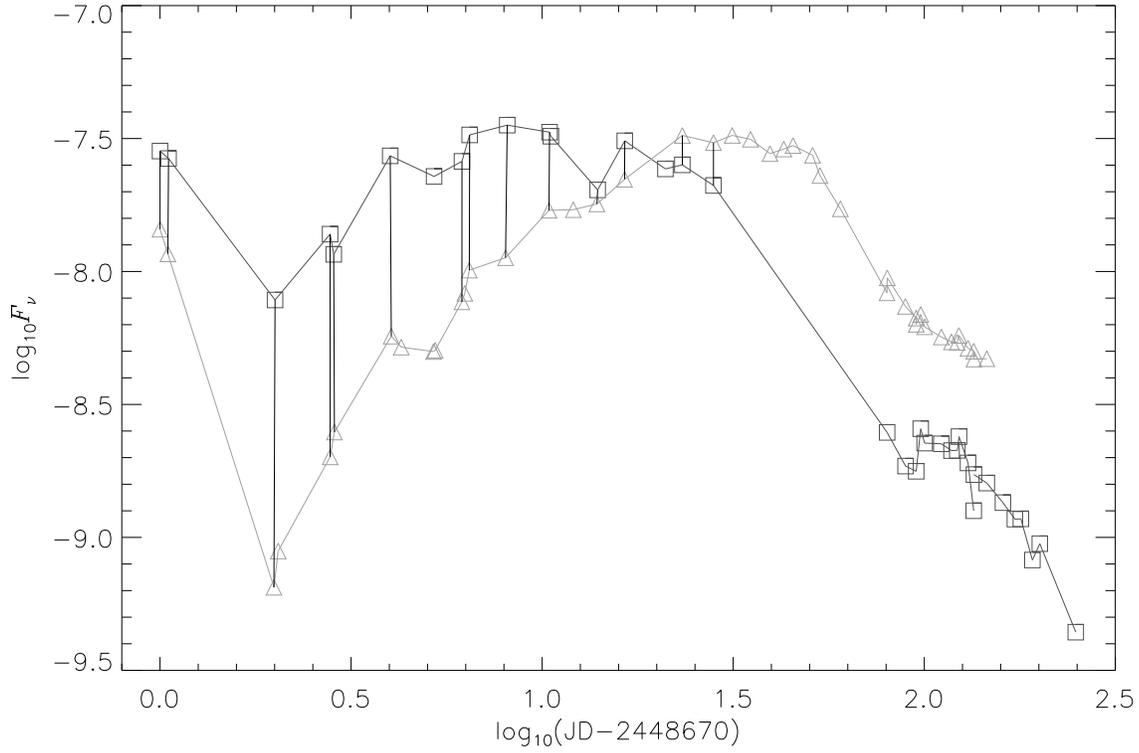}
\figcaption[f2.eps]
{UV light curves.  Mean $\lambda$ integrated flux, $\bar{F}_\lambda$, as a function of
time for the LWP (squares, darker line) and SWP (triangles, lighter line).
Vertical lines connecting the two light curves denote times of 
approximate simultaneity of the SWP and LWP observations where
IUE colors were calculated (see text). 
\label{lcurv} }
\end{figure}

   Fig. \ref{lcurv} 
shows the SWP and LWP light curves with the fourteen pairs of coincident spectra shown in 
Fig. \ref{surf_lo} marked with vertical lines.  The SWP light peaks about 20 days after the LWP light, which 
is consistent with the expected photometric time development of novae (see discussion is
Section \ref{modeling}). 
   We have formed the observed IUE color, $(SWP-LWP)_{\rm obs}$, at these times by computing the value of 
$\log (\bar{F}_{\rm LWP}/\bar{F}_{\rm SWP})$.  
The center panel of Fig. \ref{time_teff} shows the time development of $(SWP-LWP)_{\rm obs}$.  
As expected, the color becomes more positive (redder) during the 
fireball phase where $T_{\rm eff}$ is decreasing, and then gradually becomes more negative
(bluer) during the optically thick wind phase where $T_{\rm eff}$ is increasing.  

\section{Modeling}\label{modeling}

   We have calculated a set of atmospheric models that span a $T_{\rm eff}$ range
of 12 to 30 kK in increments of 1 kK.  The most important model parameters
are the
maximum expansion velocity of the linear velocity law, $v_{\rm o}$, which is 
$2000$ km s$^{-1}$ \citep{ncyg1}, the exponent of the exponential $\rho$ law, $N$, which is 3,
the micro-turbulent velocity broadening width, $\xi$, which is $50$ km s$^{-1}$, 
and the abundances, $[\frac{A}{H}]$, which are solar.  A linear velocity law and 
a value of $N$ equal to $3$ are consistent with a constant mass loss rate ($\dot{M}=$constant).  The 
radius $R=r(\tau_{5000}=1)$, where
$\tau_{5000}$ is the continuum optical depth at 5000\AA, is adjusted for each
value of $T_{\rm eff}$ to keep the bolometric luminosity, $L_{\rm bol}$, equal to 
$50\, 000 {\rm L}_\odot$.  \citet{pistin} found that the calculated spectrum is relatively
insensitive to the value of $L_{\rm bol}$, therefore, we do not vary this parameter. 
Abundance analyses of nova ejecta find them to have abundances of CNO and Fe that are
enhanced with respect to the solar values (see, for example, \citet{nebul}, or \citet{ncyg2}).
The enhanced metal abundance is thought to be caused by mixing of the ejected material
with the metal rich White Dwarf (WD) material prior to the outburst, and by  
H burning during the thermonuclear runaway (TNR) that drives a nova explosion.  
\citet{ncyg2} found $[\frac{CNO}{H}]\approx 1$ and 
$[\frac{Fe}{H}]\approx 0.3$, by number density with respect to the Sun.  Therefore, we have also 
calculated a sub-set of the grid with the hydrogen abundance reduced to half of its solar
value, by number, which makes $n_{\rm H}$ equal to $n_{\rm He}$ and $[\frac{A}{H}]$ equal to 
0.3~ (${{A/H}\over{{A/H_\odot}}}=2$ by number) for all metals.

\paragraph{}

Table 2 shows all the species that are treatable in NLTE with {\tt PHOENIX} and
indicates which ones have been included in the NLTE rate equations for each value of 
$T_{\rm eff}$ in the model grid.  Because ionization stages that are important for some values
of $T_{\rm eff}$ in the grid are unimportant for others, the species that are included in NLTE 
varies throughout the grid.  In particular, the lowest ionization stages of some species
that are included in the coolest models are excluded in hotter models, and higher
ionization stages that are included in the hottest models are excluded in the cooler 
models.  In practice, test calculations with the full suite of NLTE species were performed 
at a sparse sub-grid of models, and ionization stages that contribute less than $1\%$ of the total
population of the species at all depths were treated in LTE for nearby models in the complete grid.
Fig. \ref{surf_syn} shows the synthetic spectra in the IUE wavelength range that were 
computed with the grid of models.  The spectra were computed with a sampling, $\Delta\lambda$,
of 0.03 \AA~ so that all spectral lines are well sampled, and then were degraded to the resolution of 
the IUE low resolution grating.   

\section{Analysis}\label{analysis}

The physical interpretation of the early photometric and spectroscopic time development of
a nova in the UV is well summarized by \citet{ncyg2}. 
From Fig. \ref{surf_syn} we see that as the $T_{\rm eff}$ of the model increases, the synthetic
spectral energy distribution hardens.  This is similar to the spectral hardening with time
during the optically thick wind phase, and is opposite
to the spectral softening with time during the fireball phase shown by
the observed spectra in Fig. \ref{surf_lo}.  During the brief 
fireball phase, an initial, thin shell of gas is ejected by the blast wave that is produced by the
detonation.  This initial fireball rapidly expands and cools adiabatically
and becomes transparent after the first few days.  This accounts for the initial rapid 
decline in UV brightness.  Subsequently, we see the thicker, more extended secondary ejection
that forms an expanding photosphere around the WD.  During this optically thick wind phase, $T_{\rm eff}$
gradually increases as $\rho$ declines and the expanding atmosphere becomes 
increasingly less self-shielded from the central WD.  
At the same time,
the declining $\rho$ also causes the photospheric radius ($r(\tau=1)$) to decrease so that deeper, 
hotter layers are progressively exposed.  
As a result, the peak of the emitted flux moves to progressively shorter $\lambda$, as
can be seen from Fig. \ref{lcurv}.  
During this stage the nova evolves at constant $L_{\rm bol}$ because the expanding atmosphere 
re-processes the energy
emitted by the underlying WD, which is still burning H on its surface, and does not contain any 
energy sources of its own. 
The evolution of the UV spectrum during the optically thick wind phase can also be understood by considering
the line blanketing.  Near the beginning of the optically thick wind phase, the temperature
of the atmosphere is such that Fe is mostly in the form of \ion{Fe}{1} and \ion{}{2}, both of which have a 
very rich absorption spectrum in the UV.  As
a result, the UV flux is blocked by massive line absorption (the ``iron curtain'').  As the gas 
expands it thins and becomes increasingly radiatively heated by the central engine until it 
gradually takes on a more
nebular character and \ion{Fe}{0} becomes multiply ionized again.  As a result, the UV
opacity decreases and the UV flux increases and reaches a plateau about 50 days after the explosion.  
Some time after 150 days, the atmosphere expands to the point where the total 
optical depth through the atmosphere is less than unity and the continuum flux disappears.

\subsection{Low resolution spectra}

We have computed the synthetic IUE color, $(SWP-LWP)_{\rm syn}$, from the synthetic spectra using the same 
procedure that we used for $(SWP-LWP)_{\rm obs}$.  Fig. \ref{color} shows the relationship between 
$(SWP-LWP)_{\rm syn}$ and $T_{\rm eff}$.  As expected, the color becomes bluer as 
$T_{\rm eff}$ increases.  We note that throughout the $T_{\rm eff}$ range of
our model grid, the slope of the $T_{\rm eff}(SWP-LWP)$ relation allows for an unambiguous
$T_{\rm eff}$ determination from the $(SWP-LWP)$ color, given a sufficiently small error
margin for the color.

\paragraph{}

\begin{figure}
\plotone{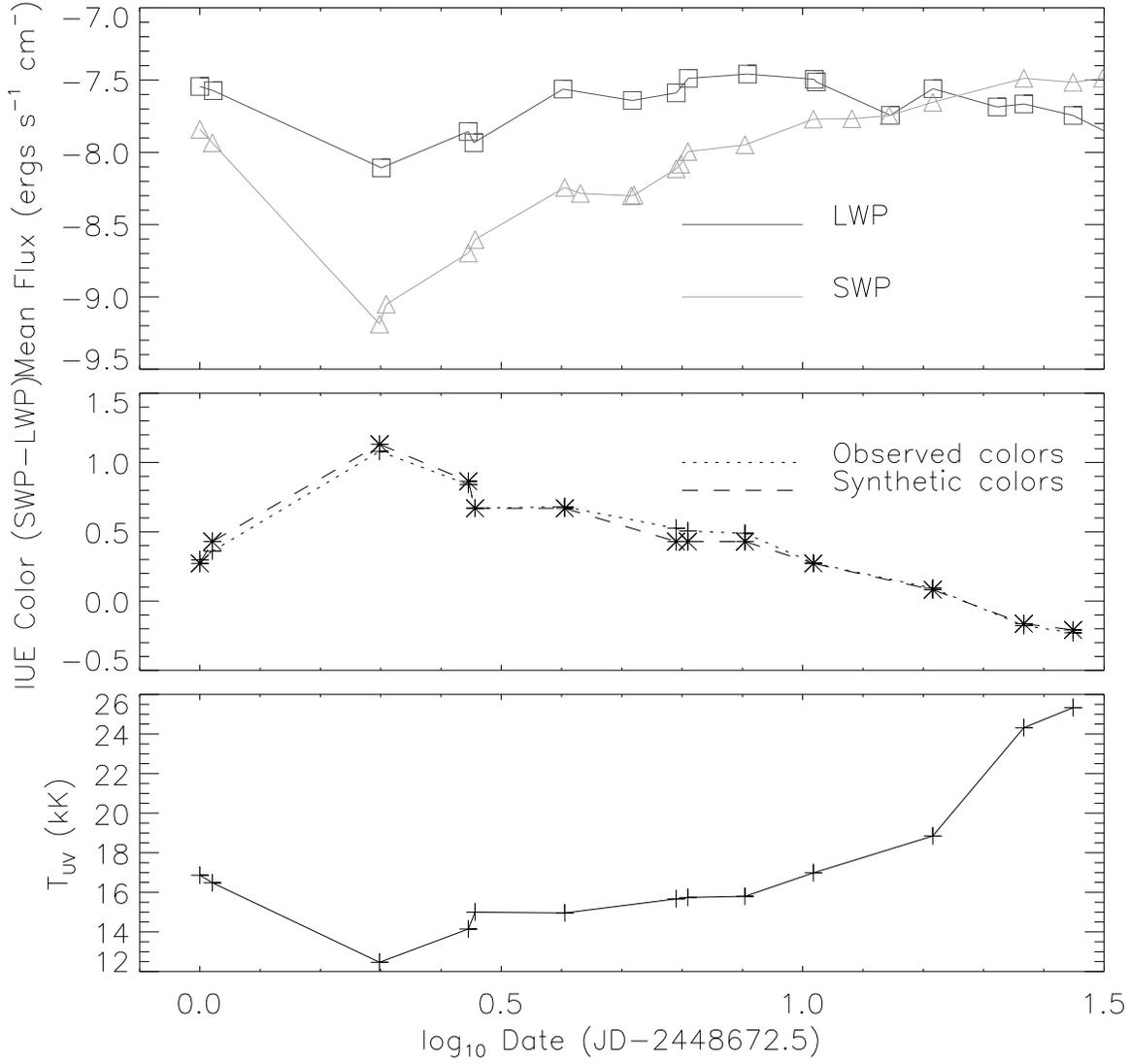}
\figcaption[f3.eps]
{Deriving the time development of $T_{\rm UV}$ from the IUE data.  Upper panel: Mean $\lambda$ 
integrated flux, $\bar{F}_\lambda$, for the LWP (squares, darker line) and SWP (triangles, lighter line).  
Center panel: IUE color, $\log\bar{F}_{\rm LWP}/\bar{F}_{\rm SWP}$.
Observed (plus symbols, dotted line), synthetic (asterisks, dashed line).  Lower panel: derived
UV color temperature, $T_{\rm UV}$ (see text). 
\label{time_teff} }
\end{figure}

\begin{figure}
\plotone{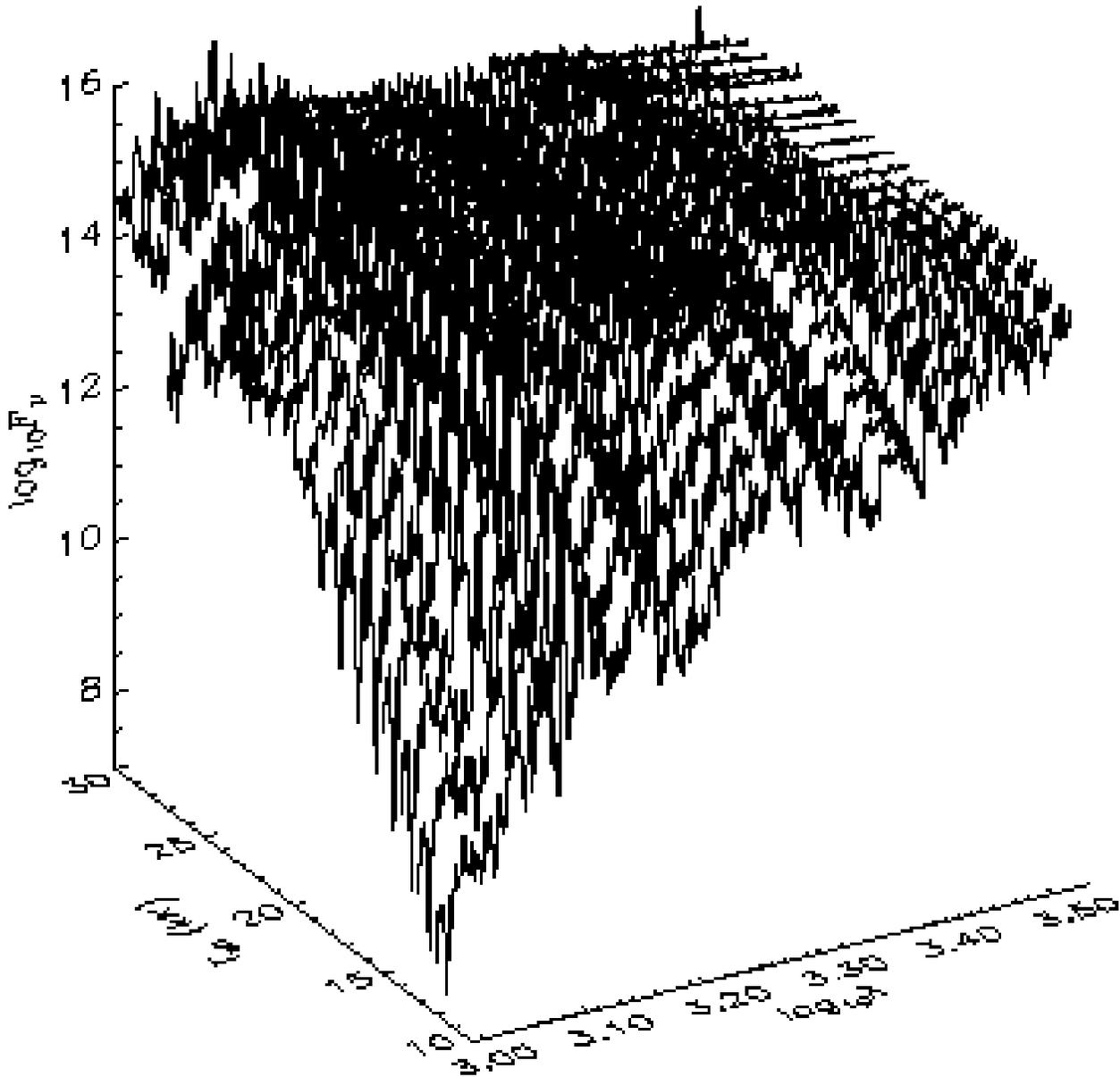}
\figcaption[f4.eps]
{Model flux surface for nova grid consists of closely spaced synthetic spectra.  Surface
gives general overview of how the global flux level and spectral energy distribution changes
with $T_{\rm eff}$.
\label{surf_syn}} 
\end{figure}

\begin{figure}
\plotone{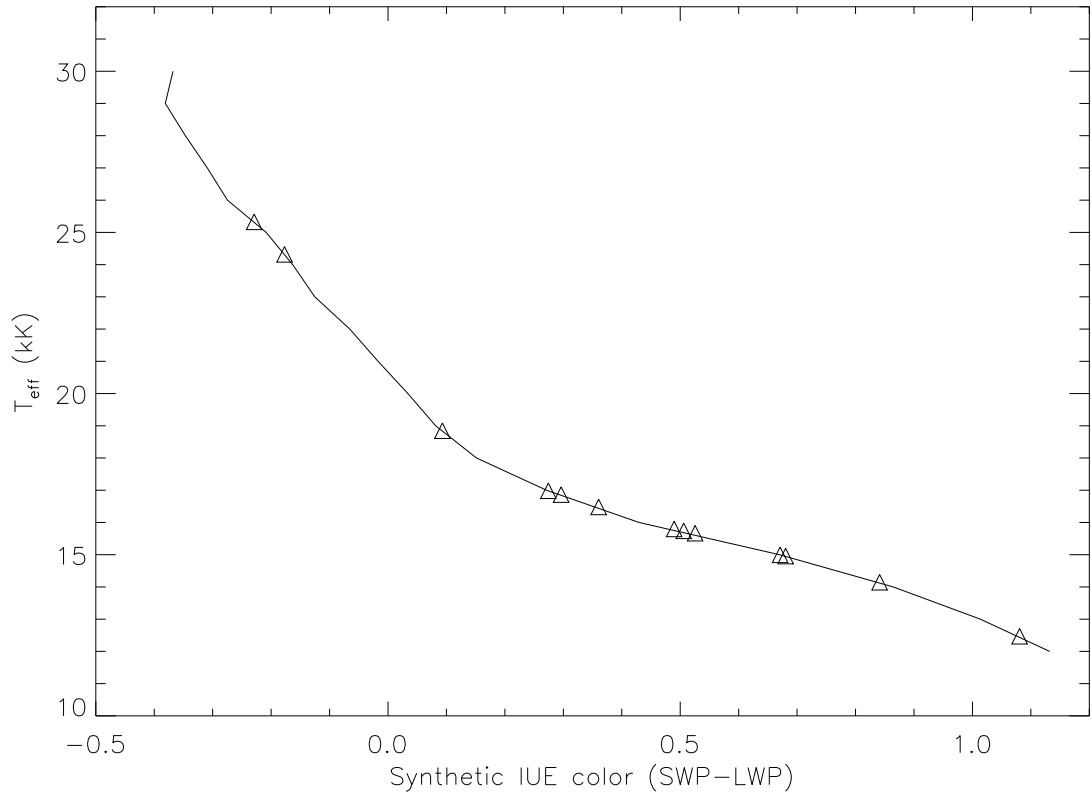}
\figcaption[f5.eps]
{Model $T_{\rm eff}$ as a function of synthetic color $(SWP-LWP)$.  
\label{color} }
\end{figure}

   We have assigned a Julian Date
to each of the synthetic colors, $(SWP-LWP)_{\rm syn}$, shown in Fig. \ref{color} by 
interpolation
within the $(SWP-LWP)_{\rm obs}(t)$ relation.  This allows us to arrange our 
models in a chronological 
sequence that reflects the time development of the nova.  The center panel of
Fig. \ref{time_teff} shows a comparison of the observed and synthetic 
color curves, $(SWP-LWP)_{\rm obs}(t)$ and $(SWP-LWP)_{\rm syn}(t)$.
We have also assigned a time sequence of UV color temperatures, $T_{\rm UV}(t)$, 
to the sequence of observed colors, $(SWP-LWP)_{\rm obs}(t)$, by interpolation within 
the $T_{\rm eff}((SWP-LWP)_{\rm syn})$ relation.
The lower panel of Fig. \ref{time_teff} shows the 
derived values of $T_{\rm UV}$ as a function of time.  The
UV radiation temperature in the IUE range initially cools from $\sim17$ kK to 
$\sim12$ kK during the fireball phase,
then gradually heats up during the optically thick wind phase, till it reaches
$\sim24$ kK about 30 days after the explosion.  

\paragraph{}

  Because the IUE $\lambda$ range is close to the peak of the spectral
energy distribution, $F_\lambda(\lambda)$, 
for this $T_{\rm eff}$ range, we have approximated the time development 
of $T_{\rm eff}$, $T_{\rm eff}(t)$,
by setting it equal to $T_{\rm UV}(t)$.  On this basis, we assign a synthetic
spectrum from the model grid to each of the times marked in Fig. \ref{lcurv}.

\paragraph{}

\begin{figure}
\plotone{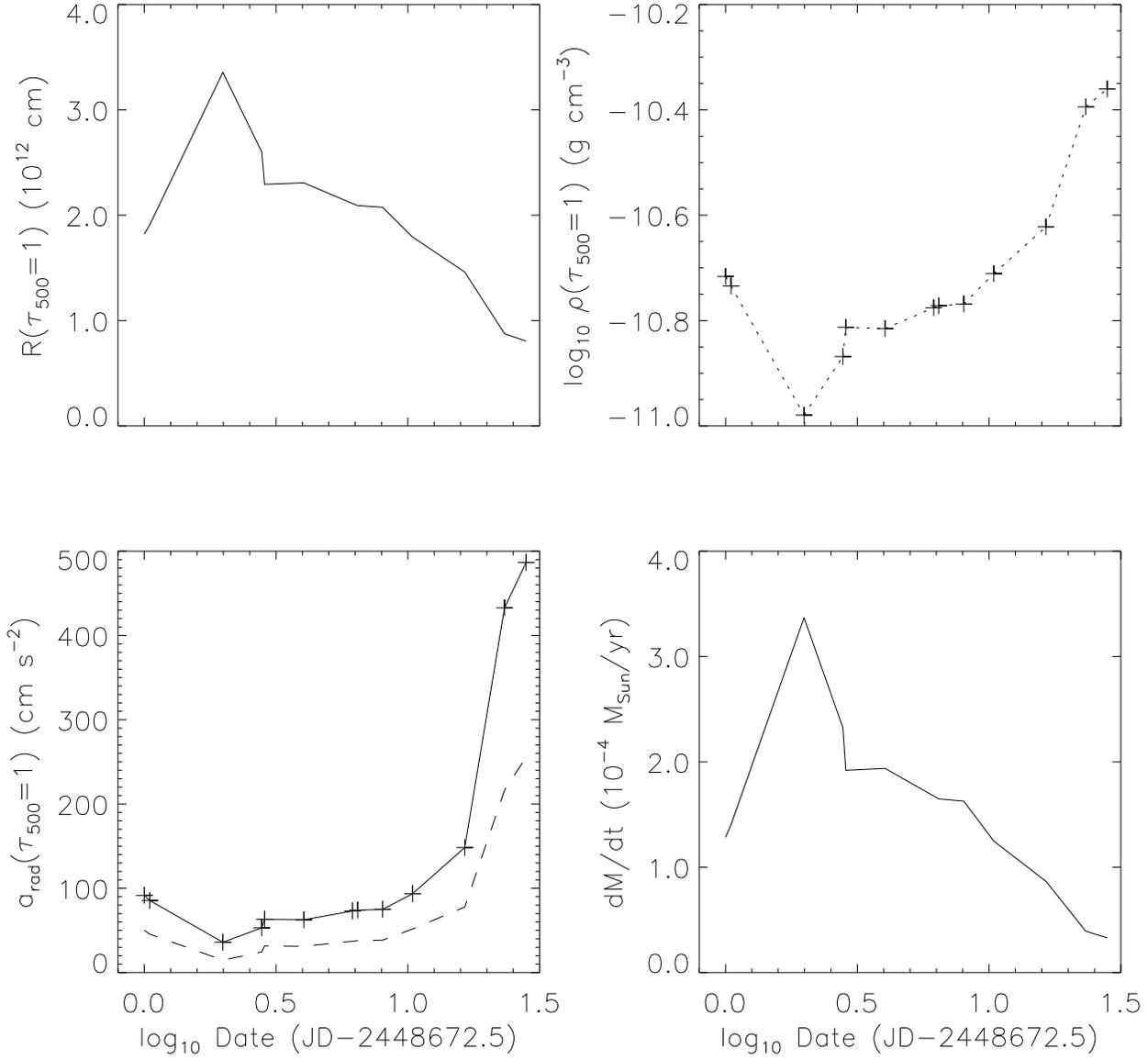}
\figcaption[f5a.eps]
{The time development of four model quantities according to the 
chronological ordering based on the IUE color (see text).  Upper 
left panel: $R(\tau_{\rm 5000}=1)$; upper right panel: 
$\log\rho(\tau_{\rm 5000}=1)$; lower left panel:
$|a_{\rm rad}(\tau_{\rm 5000}=1)|$, and, for reference,
the value of $g$ assuming a 1.25 $M_\odot$ WD; lower right panel: $dM/dt$. 
\label{timedev} }
\end{figure}

Fig. \ref{timedev} shows the time development of four model quantities,
according to the chronological ordering established above: $R(\tau_{5000}=1)$, 
$\log\rho(\tau_{5000}=1)$, the {\it absolute value} of the outward acceleration due to radiation pressure
at a continuum optical depth of unity, 
$|a_{\rm rad}(\tau_{5000}=1)|$, and the total mass loss rate per year, $dM/dt$.
As discussed above, the value of $R(\tau_{5000}=1)$ decreases throughout
the optically thick wind phase because, although the atmosphere is expanding,
the physical radius at which is becomes optically thick (the photosphere) is contracting as
the gas becomes thinner.  The recession of the photosphere to deeper 
atmospheric layers also causes
the value of $\rho(\tau_{5000}=1)$  to increases
during this time.  The value of $|a_{\rm rad}(\tau_{5000}=1)|$ increases
during this time because the radiation pressure increases as the atmosphere 
becomes hotter.  For reference we have also plotted $g$, the inward
acceleration due to gravity at $R(\tau_{5000}=1)$, assuming a $1.25 M_\odot$
WD \citep{wdmass}.  The radiative acceleration of the atmosphere is discussed
further below.  
The value of $g$ increases with time as $R(\tau_{5000}=1)$ decreases,
but is always {\it less} that the value of $|a_{\rm rad}(\tau_{5000}=1)|$. 
At each time, the value of $dM/dt$ is independent of 
depth in the atmosphere, which is consistent with a linear velocity law, but 
{\it not} with a 
radiatively driven velocity law.  The value of $dM/dt$ increases with
time as the value of $|a_{\rm rad}|$ increases.

\paragraph{}

\begin{figure}
\plotone{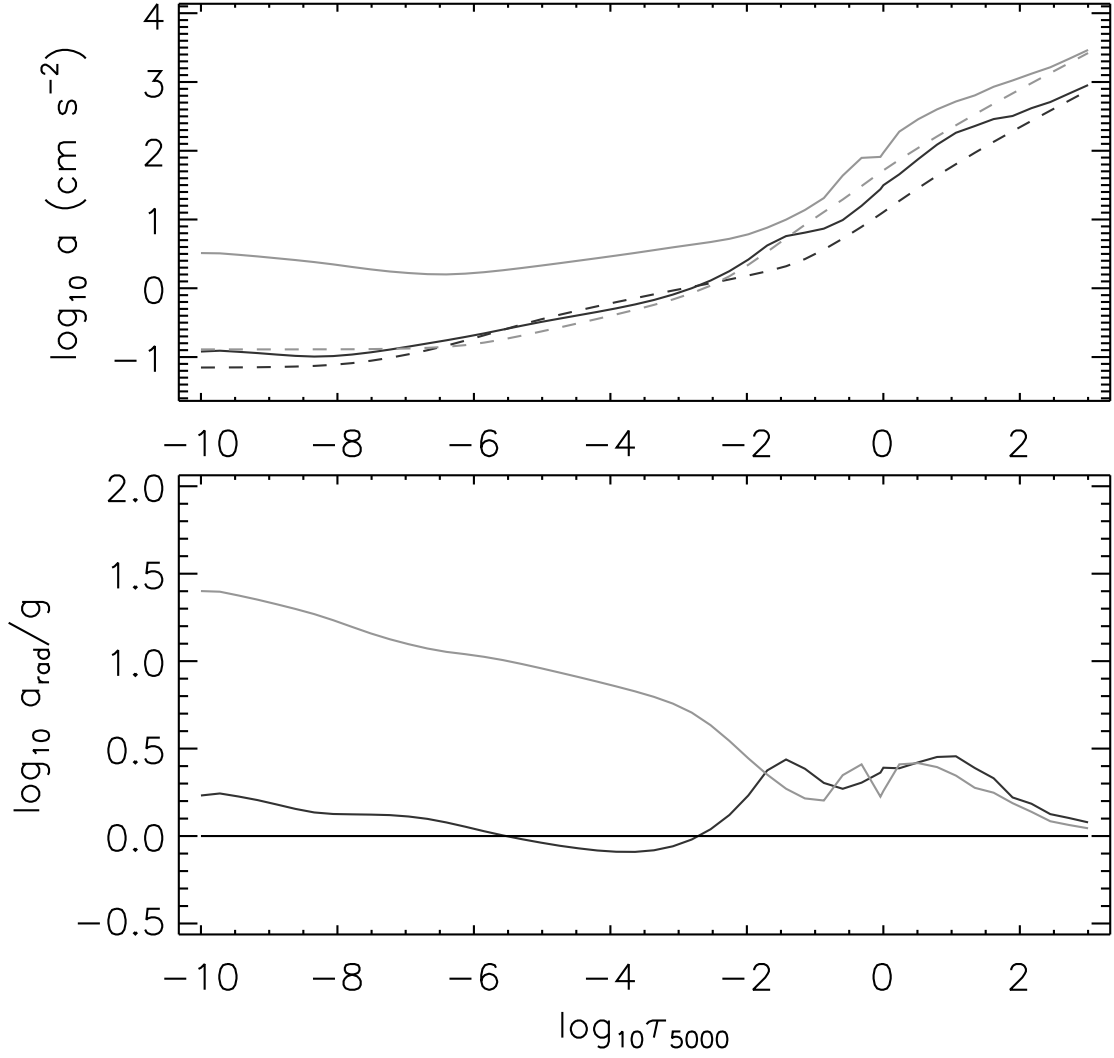}
\figcaption[f5b.eps]
{Comparison of $\log |a_{\rm rad}|$ and $\log g$ throughout two models 
that represent
two different times during the optically thick wind phase.  Darker line:
JD 2448674.40320; lighter line: JD 2448688.85069.  Upper panel: 
$\log |a_{\rm rad}|$ (solid line); $\log g$ (dashed line).  Lower panel:
$\log |a_{\rm rad}|/g$.
\label{arad}} 
\end{figure}

  Fig. \ref{arad} shows a comparison of $\log |a_{\rm rad}|$ and
$\log g$ as a function of depth throughout two of the models: 1) a model
of $T_{\rm eff}=12$ kK, which corresponds to the time of the first spectrum 
of the optically thick wind phase (JD 2448674.40320), and 2) a model of $T_{\rm eff}=17$ 
kK, which corresponds to a time about two weeks later later (JD 2448688.85069).  
The figure
also shows the difference of these quantities (ie $\log |a_{\rm rad}|/g$).
For the cooler model, $|a_{\rm rad}|$ is greater than $g$ by as much
as a factor of three at depths where $\log\tau_{\rm 5000}$ ranges from
$2$ to $-2$. 
It is slightly less than $g$ in the $\log\tau_{\rm 5000}$ range between 
$-2.5$ and $-6$,
and exceeds $g$ again in the outermost part of the atmosphere.
The hotter model shows similar behavior in the $\log\tau_{\rm 5000}$ range 
between $2$ and $-2$, but $|a_{\rm rad}|$ remains greater than $g$ everywhere.
These results suggest that radiation pressure could possibly play a role
in determining the velocity structure of the wind.  We are currently
implementing the treatment of dynamics in {\tt PHOENIX} with the goal
of investigating this question further.  

\paragraph{}

\begin{figure}
\plotone{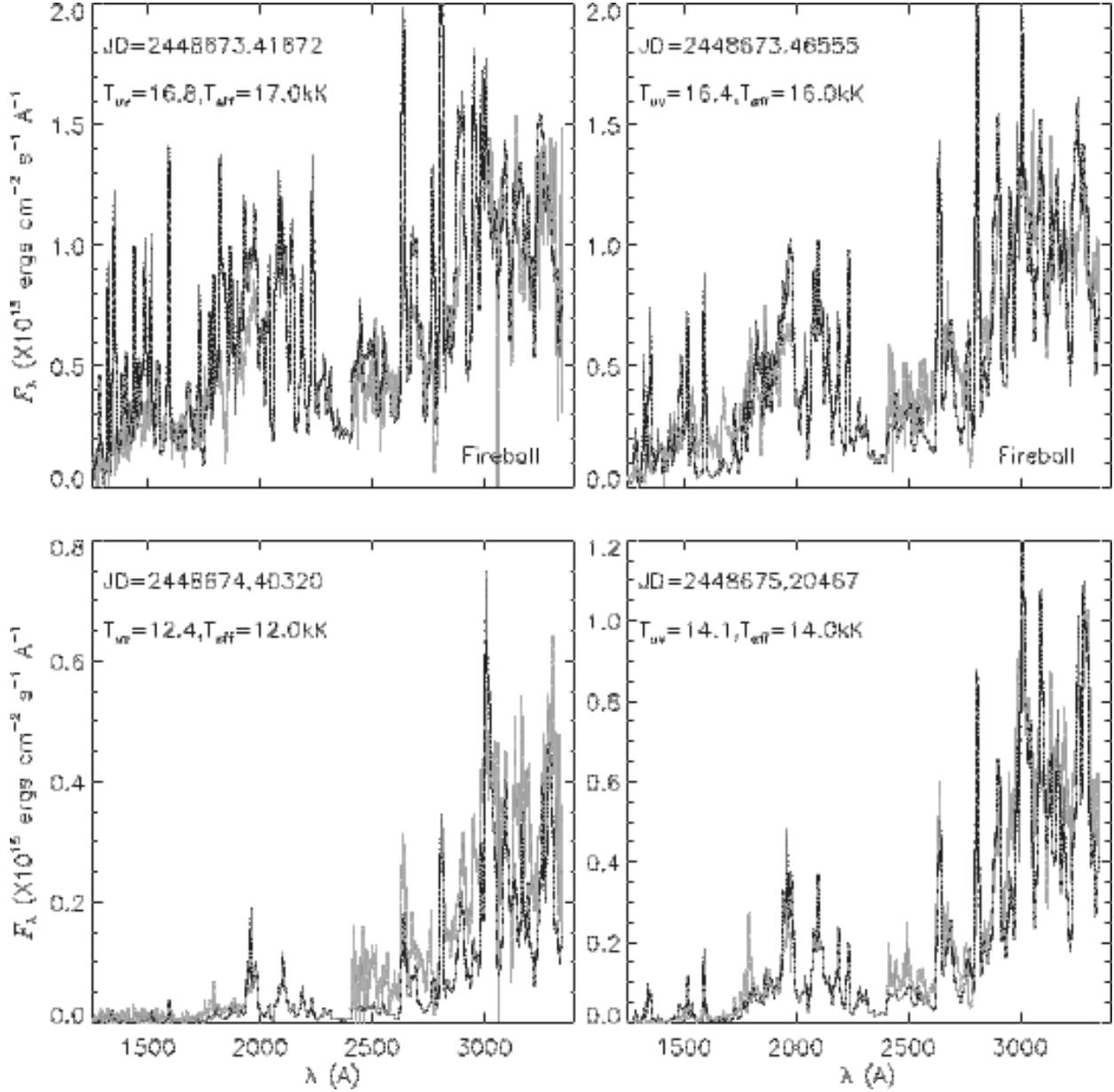}
\figcaption[f6.eps]
{Comparison of observed (lighter line) and synthetic (darker line) spectra
at the first twelve phases of approximate band simultaneity.  The gap
in $\lambda$ for the observed spectra occurs because the reduced SWP and LWP 
spectra are disjoint in $\lambda$.  Each panel is
labeled with the JD of the observed spectrum, the derived value of $T_{\rm UV}$
from the $(SWP-LWP)$ color, and the value of $T_{\rm eff}$ of the model used
to synthesize the spectrum. 
\label{temp_plot_0} }
\end{figure}

\begin{figure}
\plotone{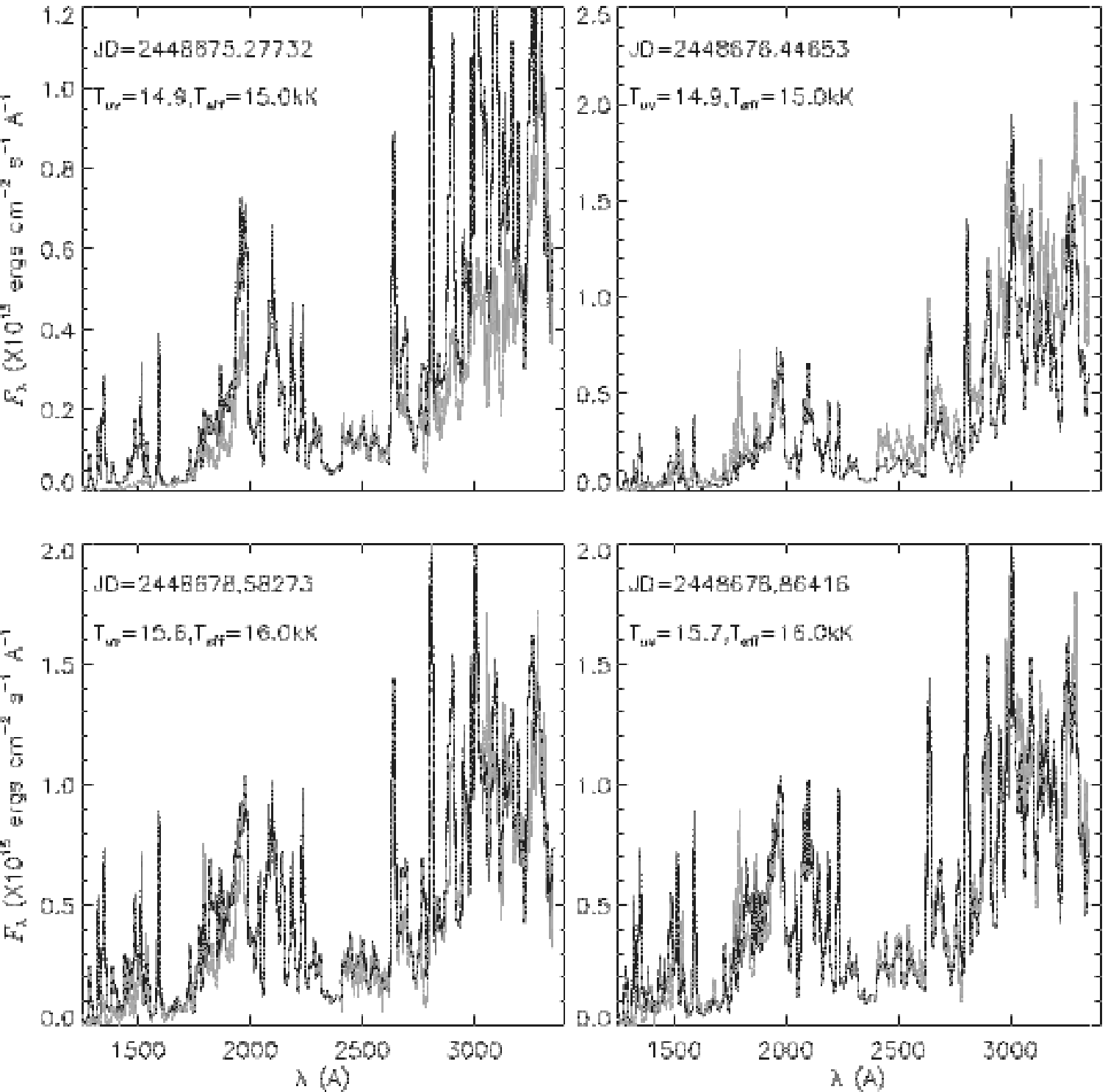}
\figcaption[f7.eps]
{See Fig. \ref{temp_plot_0}
\label{temp_plot_4} } 
\end{figure}

\begin{figure}
\plotone{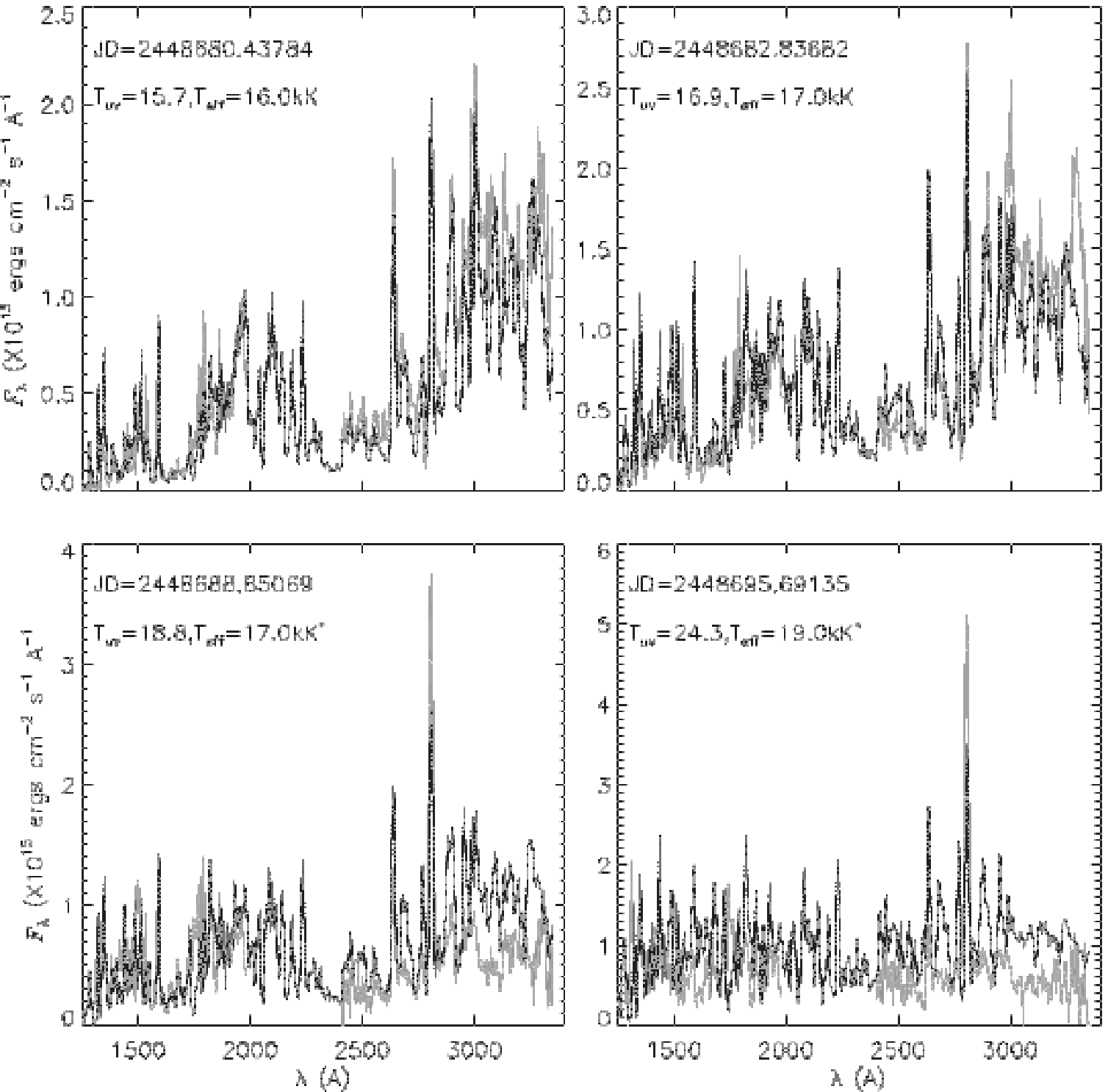}
\figcaption[f8.eps]
{See Fig. \ref{temp_plot_0}
\label{temp_plot_8} }
\end{figure}

Figs. \ref{temp_plot_0} through \ref{temp_plot_8} show for each of these times
the comparison between the observed spectrum and the synthetic spectrum for the 
model whose $T_{\rm eff}$ value is closest to the $T_{\rm UV}$ value of the 
observed spectrum.  Because we do not know the angular diameter of Cygni 92, we
arbitrarily adjust the flux level of the observed spectra to approximately match the
absolute flux level of the synthetic spectra.  However, note that in keeping with the
condition of constant $L_{\rm bol}$, the {\it same} scale 
factor has been used in all plots ({\it ie.} individual pairs of observed and
synthetic spectra have {\it not} had their flux calibration ``tuned'').  The observed 
spectrum in each
panel has a gap in the middle because the reduced SWP and LWP spectra are 
disjoint in $\lambda$.  For times later than JD 2448683, the \ion{Mg}{2} $hk$
resonance doublet at $\lambda 2800$ becomes increasingly strong in emission with
respect to the pseudo-continuum, which indicates that the nova is becoming increasingly
nebular at that time.  

\paragraph{}
  
   For the fireball phase, and for all times throughout the optically thick wind phase
up to JD 2448682.8368, 
a synthetic spectrum of $T_{\rm eff}\approx T_{\rm UV}$ provides 
a good fit to the global shape of the pseudo-continuum throughout the entire 
IUE $\lambda$ range.  This is significant because we 
see from both Figs. \ref{surf_lo} and \ref{surf_syn} that the SWP range is 
sensitive to $T_{\rm eff}$ in this $T_{\rm eff}$ range.  

\paragraph{}

   To illustrate the similarity of the fireball phase to the 
optically thick wind phase, we have identified two times, JD2448673.4167
(fireball phase), and JD2448686.3241 (optically thick wind 
phase) at which the $(SWP-LWP)_{\rm obs}$ color, and, therefore, the value of $T_{\rm UV}$,
is approximately the same.  Fig. \ref{fireball2} shows a comparison of the IUE
spectra at these two times.  Both spectra have been multiplied by the {\it same} factor to 
approximately remove the dilution factor, but they have {\it not} been shifted with 
respect to one another.  The absolute flux level and overall shape of the
pseudo-continuum is approximately the same.  However, the optically thick wind phase spectrum 
exhibits significantly stronger emission lines than the fireball phase spectrum.  The difference 
in the strength of the emission lines is due to the greater amount of
Doppler broadening from a steeper velocity gradient through the LFR in the thin fireball
\citep{ncyg2}.

\paragraph{}

\begin{figure}
\plotone{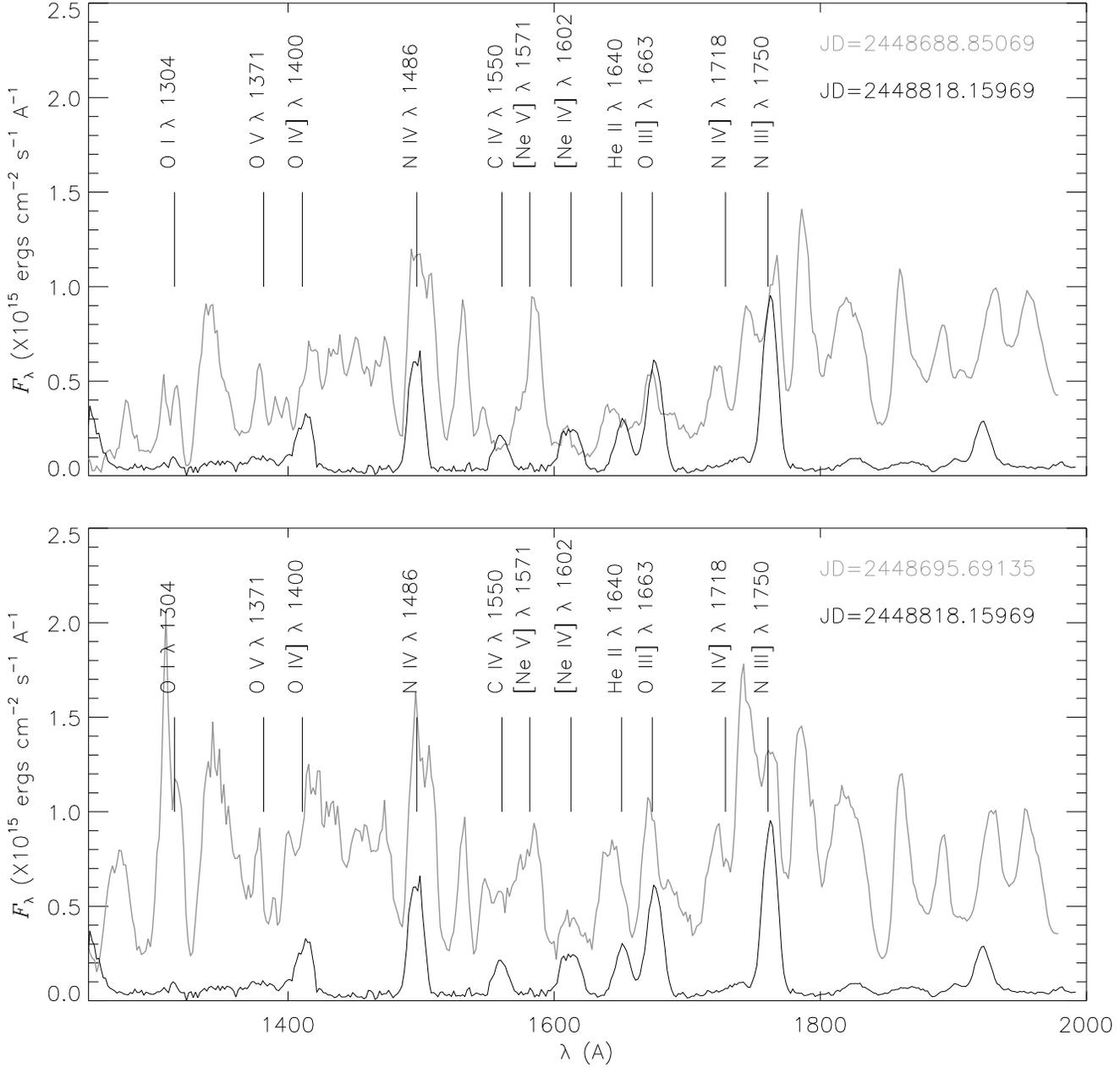}
\figcaption[f9.eps]
{The last two observed spectra that were fit in Fig. \ref{temp_plot_8} (JD2448688.85069
and JD2448695.69135) (lighter line).  A spectrum from the nebular phase of the outburst taken
over 100 days later (darker line).  The line annotations refer to nebular lines. 
\label{nebular}}
\end{figure}

\begin{figure}
\plotone{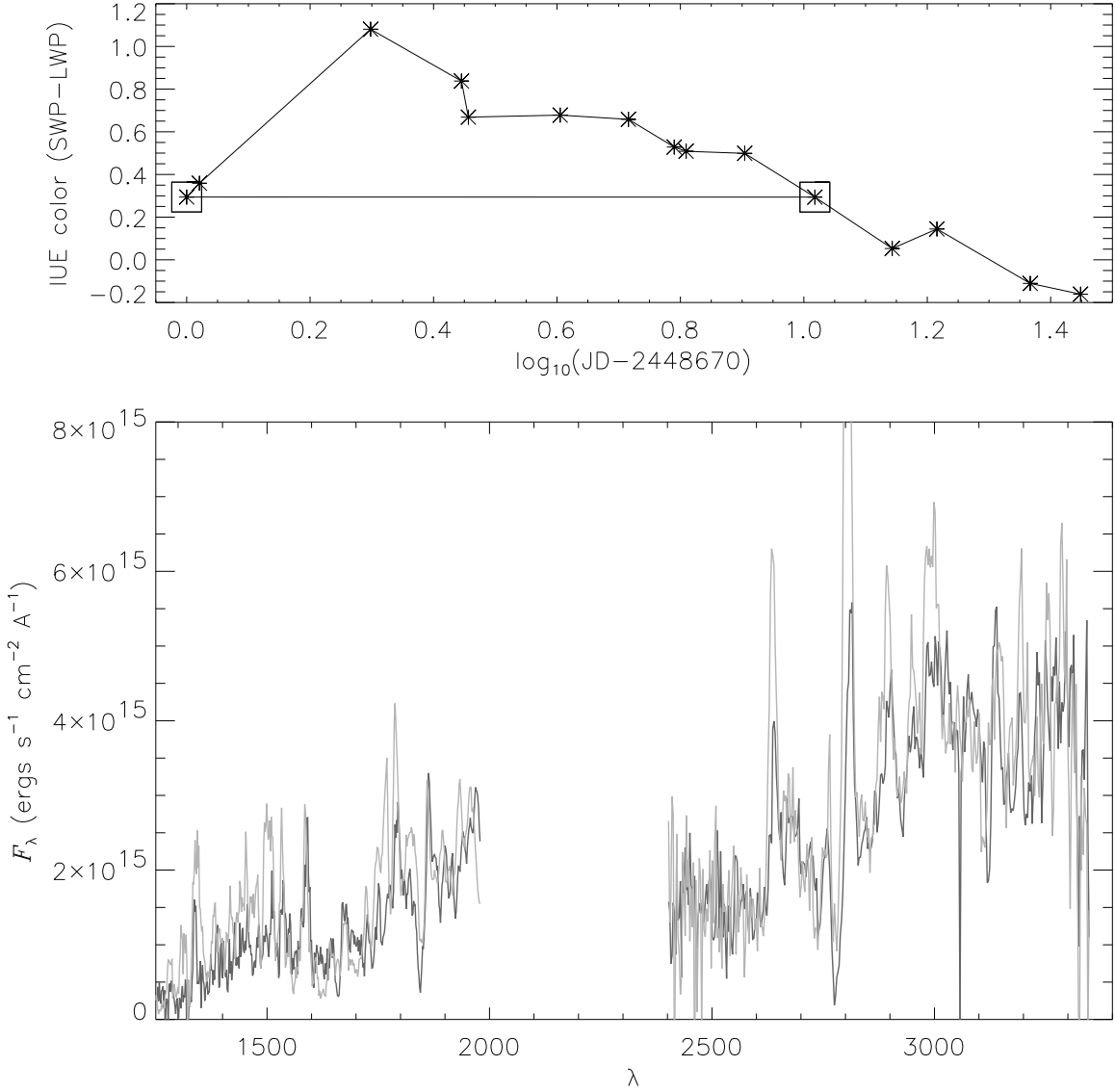}
\figcaption[f10.eps]
{Comparison of fireball and optically thick wind phase at equal color.  Upper panel: observed
IUE color $(SWP-LWP)$ as a function of time.  The squares connected by the horizontal line indicate
the two phases being compared in the lower panel.  Lower panel: Comparison of fireball spectrum 
(darker line) and optically thick wind phase spectrum (lighter line) at the two phases indicated
in the upper panel.
\label{fireball2} }
\end{figure}

For the last two times shown, JD2448688.8507 and JD2448695.6914, none of the
synthetic spectra provide as close a fit as that for earlier times.  Furthermore,
in contrast to what we find for earlier times, the closest fit synthetic spectrum 
is that of a model for which $T_{\rm eff}$ is significantly lower that the
observed value of $T_{\rm UV}$.  Fig. \ref{nebular} shows the observed spectrum
at these two times compared with an IUE low resolution spectrum that was taken
over 100 days later (SWP45135).  At the later time, the nova has become nebular and exhibits a
pure emission spectrum with permitted, semi-forbidden, and forbidden lines in the UV.  
The velocity of the outflowing gas during the nebular phase is $4500$ km s$^{-1}$
(\citep{nebul}).  Therefore, we have shifted the nebular spectrum red-ward by
$2000$ km s$^{-1}$.  The nebular emission lines labeled in Fig. \ref{nebular} were
identified using Table 1 in \citet{nebul}.  The comparison of the nebular phase
spectrum with the two earlier spectra shows that by JD2448688.8507 many of the nebular 
lines were already present in the spectrum.  The expanding atmosphere was
already partially nebular by JD2448688.8507 in that many of the nebular lines had already
begun to ``contaminate'' the optically thick wind phase spectrum.  As a result,
we expect that {\tt PHOENIX} models will provide an increasingly worse fit beyond
this time because {\tt PHOENIX} is capable of modeling only the component
of the outburst that constitutes an optically thick atmosphere.

%

\subsection{High resolution spectra}

\begin{figure}
\plotone{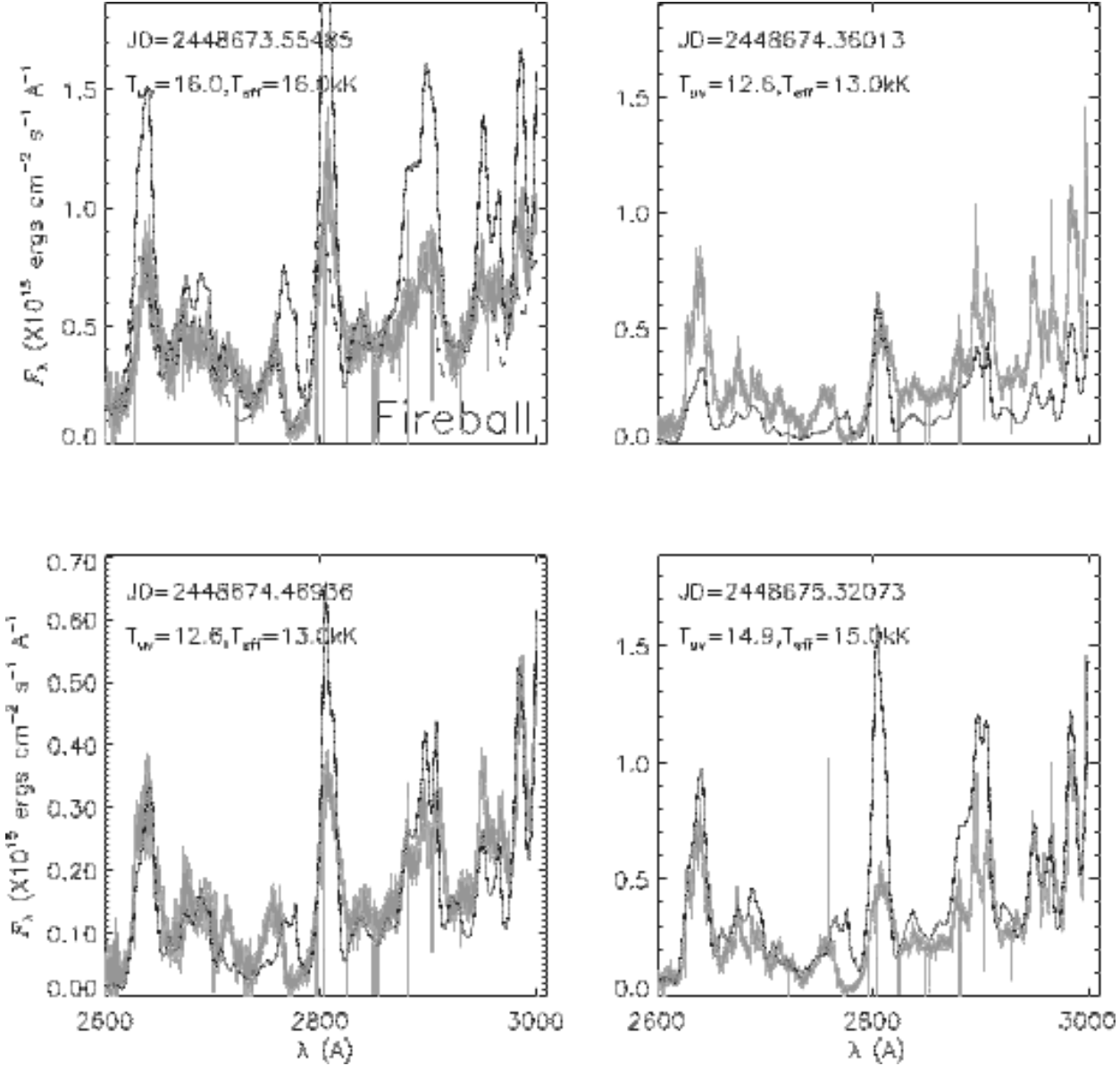}
\figcaption[f11.eps]
{Comparison of high resolution observed (lighter line) and synthetic (darker line) spectra
at eight phases.  Each panel is
labeled with the JD of the observed spectrum, the interpolated value of $T_{\rm UV}$
from the time vs $(SWP-LWP)$ relation (see text), and the value of $T_{\rm eff}$ of the model used
to synthesize the spectrum.  Upper left panel: the dashed line is the synthetic 
spectrum of a supernova model with $T_{\rm eff}$ equal to 12 kK.
\label{hi_plot0} } 
\end{figure}

\begin{figure}
\plotone{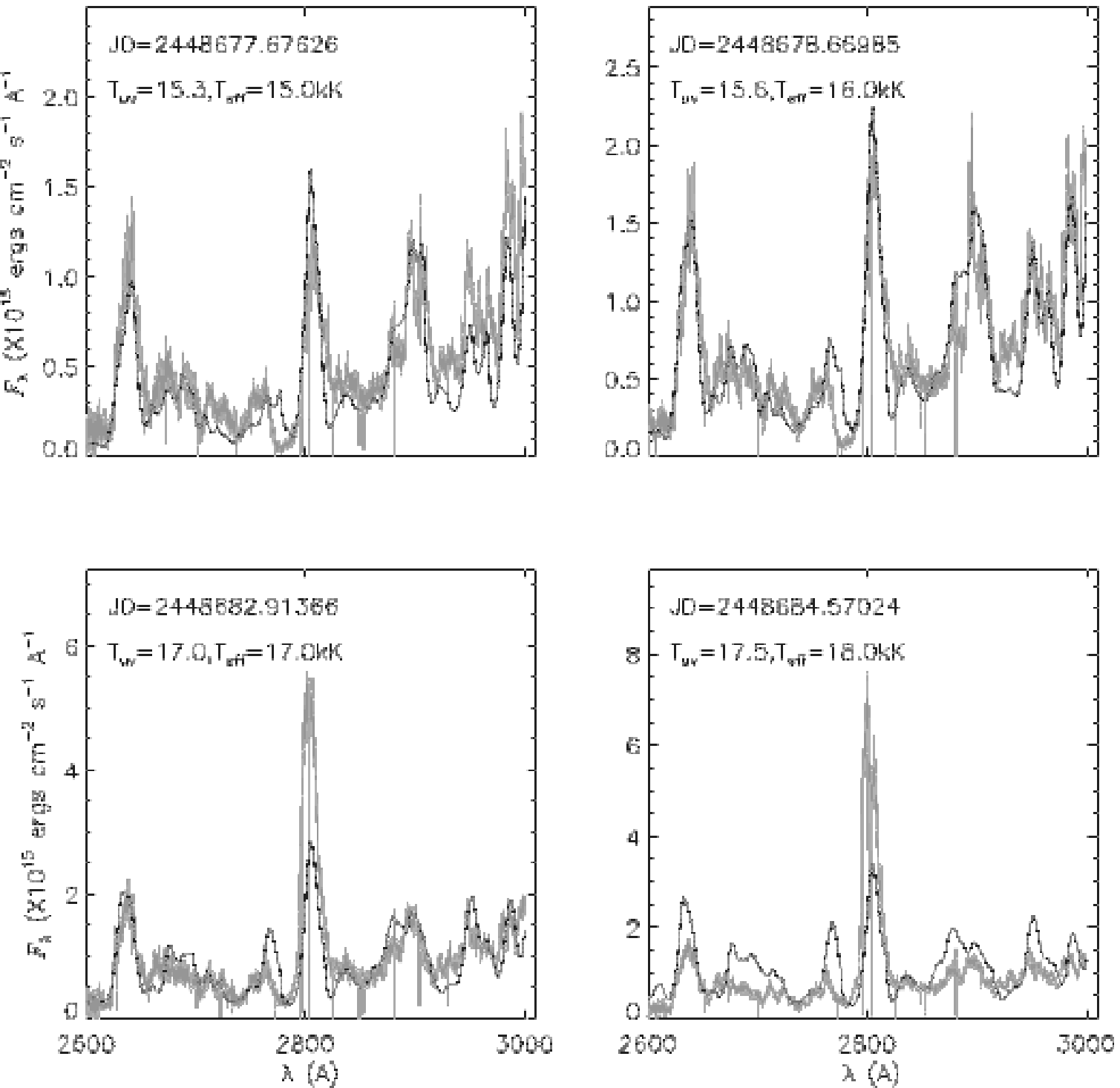}
\figcaption[f12.eps]
{See Fig. \ref{hi_plot0}
\label{hi_plot4} } 
\end{figure}

    Figs. \ref{hi_plot0} and \ref{hi_plot4} show a time sequence of high resolution IUE spectra in the
LWP pass band that overlaps in time the sequence shown in Figs. \ref{temp_plot_0} through \ref{temp_plot_8}.
Synthetic spectra have been assigned to each time by interpolation in the time vs $T_{\rm UV}$ 
relation that was derived from the low resolution spectra (lower panel of 
\ref{time_teff}).   The prominent emission feature at $\lambda 2800$ is the $hk$ resonance doublet
of \ion{Mg}{2}.  The \ion{Mg}{2} $hk$ lines become stronger with time because the expanding
atmosphere gradually becomes more nebular.  As a result, the models, which only account for 
the photospheric part of the expanding gas shell, increasingly underpredict the strength of these 
lines as time increases.  The prominent emission
features at $\lambda 2630$ and $\lambda 2900$, which are approximately reproduced by the models at many 
times, are not emission lines, but
are regions of relatively lower line opacity through which more flux escapes than through the neighboring 
regions.  

\paragraph{}

Generally, the models approximately reproduce the broad features of the high resolution spectra
during the optically thick wind phase.  However, there are discrepancies in the detailed fit.
One possible source of discrepancy is the distribution of metal abundances.  The nova ejecta
may be depleted in H and enriched in C, N, and O as a result of WD material having been 
mixed into the accreted layers, or as a result of nuclear processed elements from the surface H
burning that drives the nova explosion having been ejected.  \citet{ncyg2} found that $[\frac{CNO}{H}]$
is approximately one while $[\frac{Fe}{H}]$ is only about 0.3.  Some discrepancy may also
be due to the form of the velocity law.  If the ejecta are being driven outward by radiation pressure
at this stage, then $v(r)$ may have the power law form of a wind rather than a linear form \citep{CAK}.  Indeed, since the time of \citet{mclaugh} it has been 
observed that relatively blue shifted spectral absorption components arise
in relatively deep layers of the expanding atmosphere, which suggests a
more complex velocity law than the linear one employed here.  
We are currently in the process of converging power law wind models. 
 Another possible source of
discrepancy is inhomogeneities in the expanding gas shell, which have been observed (see, for example,
\citet{payne} or \citet{nebul}), and which are expected to arise if the velocity
field is complex.  These will serve to make the line
profiles more complex than those predicted with a homogeneous model.

\paragraph{}

The upper left panel in Fig. \ref{hi_plot0} shows the fit to the only high resolution spectrum
taken during the fireball phase.  At this resolution it is clear that the nova models systematically
over-predict the strength of all emission features by a factor of approximately two.  The dashed
line is the synthetic spectrum of a model with $T_{\rm eff}$ equal to $12000$ K, $v_{\rm max}$
equal to $4000$ km s$^{-1}$, and $N=7$ rather than $3$ ($\rho(r)~\alpha ~r^{-N}$).  
This model is much thinner in the radial direction and has much steeper $\rho(r)$ and $v(r)$ gradients
than the models in our nova grid.  Moreover, because $N>3$, the motion of the atmosphere does
not correspond to a constant mass loss rate, but, rather to homologous expansion.  The
characteristics of this model are more 
similar to those of supernova models rather than typical nova models.  Such a model yields a synthetic
spectrum with much weaker emission features than that of a more typical nova model because the steeper 
$v(r)$ gradient leads to a greater Doppler smearing of the emergent spectrum.  As a
result, this model provides a significantly better fit to the observed spectrum.  This is
consistent with the idea that the fireball phase is due to a thin shell of gas that is ejected
by the blast wave before the expanding atmosphere that causes the optically thick wind phase
develops.  We note
that the best fit value of $T_{\rm eff}$ for the fireball spectrum is $4000$ K lower when it
is fit with the thinner, steeper model rather than a typical nova model.
\citet{ncyg2} found that a thin shell model with a steep ($N=15$) $\rho$ law
and enhanced \ion{CNO}{0} and \ion{Fe}{0} abundances was required to fit
the first two observed spectra.  A comparison of the 2600 to 3000 \AA~ region in their Fig. 5b with 
Fig. \ref{hi_plot0} shows
that we achieve a better fit to the fireball phase with a solar abundance model than they did. 
The improvement may be due to the greater number of species treated in NLTE.  

\paragraph{}
 
\begin{figure}
\plotone{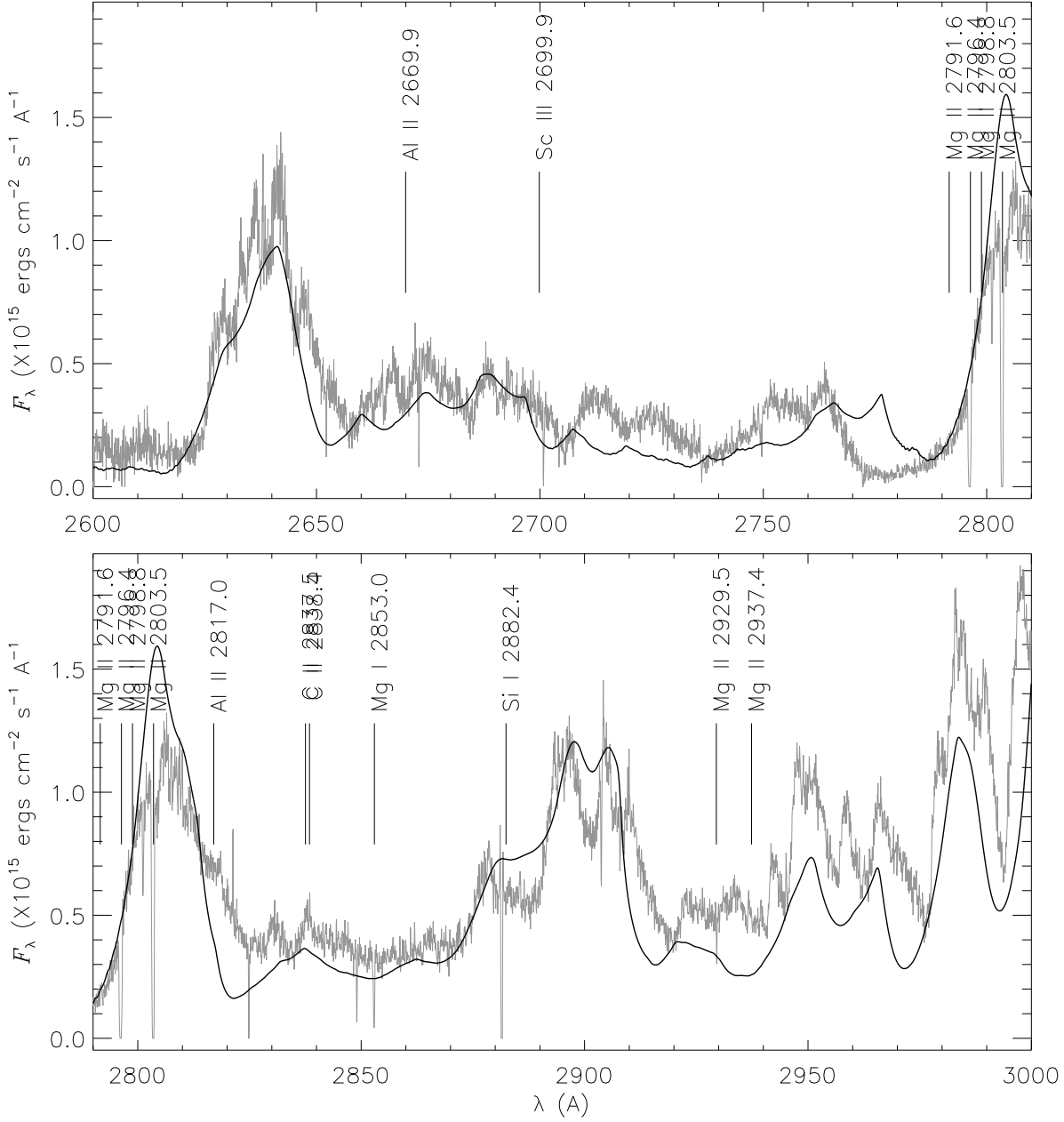}
\figcaption[f13.eps]
{Comparison of high resolution observed (lighter line) and synthetic (darker line) spectra
at a representative time during the optically thick wind phase.  
The strongest lines in the spectrum synthesis, other than Fe group lines, have been labeled with
their identifications.
\label{hires_id} }
\end{figure}

    Fig. \ref{hires_id} shows a representative high resolution observed spectrum from the early optically thick
wind phase (LWP22457, JD2448677.67626, $T_{\rm UV}\approx 15$ kK).  Also shown is the best fit
synthetic spectrum.  The plot has been labeled with the identifications of the strongest lines,
as determined by the spectrum synthesis calculation.  During the line identification procedure the Fe group
elements were ignored to prevent the plot from being completely saturated with the labels of \ion{Fe}{2}
and, to a lesser extent,
of \ion{Fe}{1}, \ion{Fe}{3}, \ion{Ti}{2}, \ion{Cr}{2}, \ion{Co}{2}, and \ion{Ni}{2}, which are the most 
important source of opacity at most wavelength points.

\subsection{Abundance effects}

\begin{figure}
\plotone{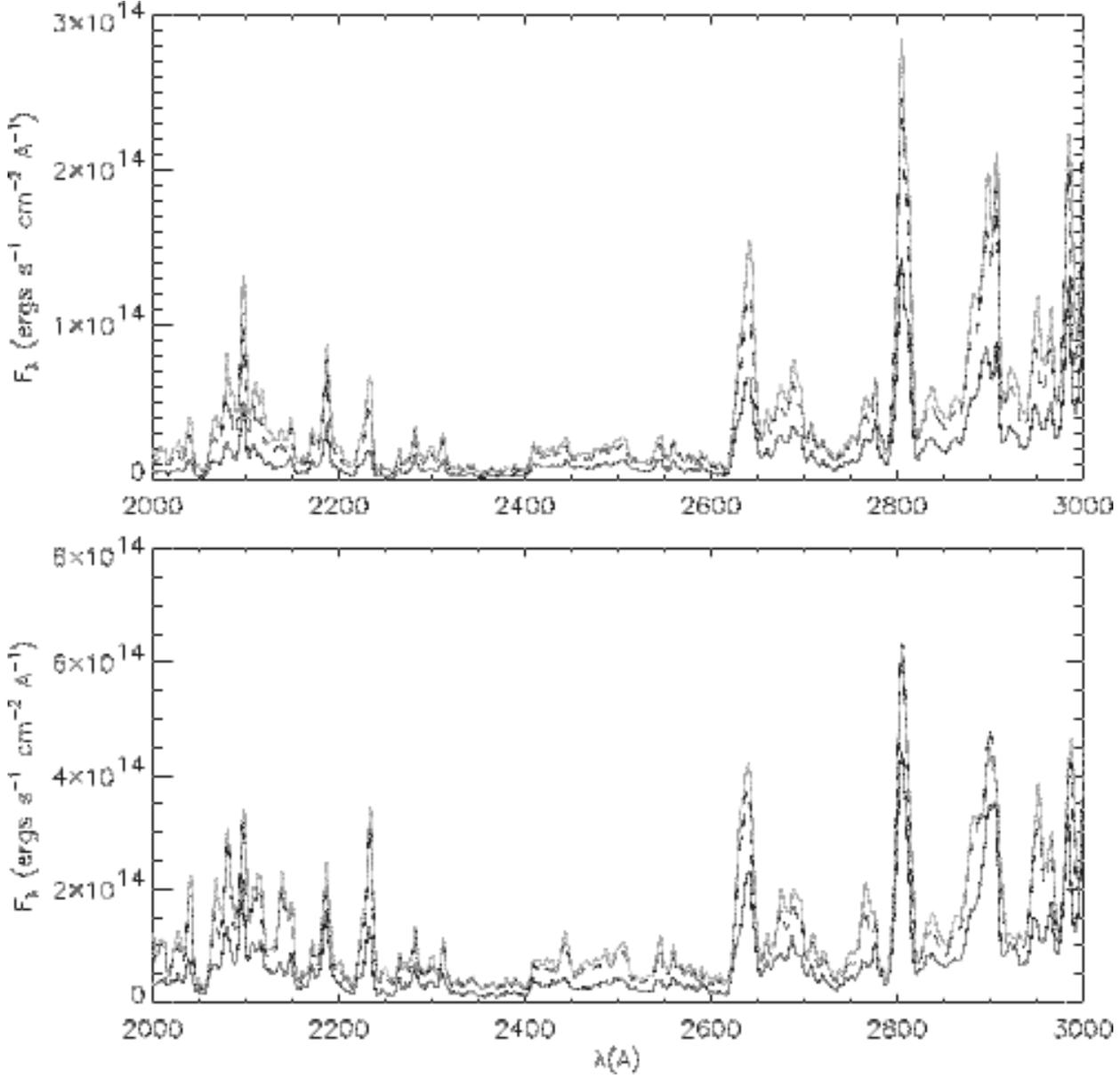}
\figcaption[f14.eps]
{Comparison of synthetic spectra with various values of $T_{\rm eff}$ and $[\frac{A}{H}]$.
Lighter line: Solar abundances, $T_{\rm eff}=14$ kK (upper panel) and $16$ kK (lower panel).
Darker solid line: $[\frac{A}{H}]=0.3$, $T_{\rm eff}=14$ kK (upper panel) and $16$ kK (lower panel). 
Darker dashed line: $[\frac{A}{H}]=0.3$, $T_{\rm eff}=15$ kK (upper panel) and $17$ kK (lower panel). 
\label{2H} }
\end{figure}

  Fig. \ref{2H} shows a comparison between the synthetic spectra of models with solar abundance and 
those that are hydrogen deficient in a representative region of the IUE spectral range.  Reducing 
$n_{\rm H}$ to half of its solar value by number makes 
$\frac{n_{\rm He}}{n_{\rm H}}=1$, and $[\frac{A}{H}]=0.3~ ({{A/H}\over{{A/H}_\odot}}=2$ 
by number) for all metals.  
From Fig. \ref{2H} we see that the hydrogen deficiency suppresses the
overall flux throughout the near UV by approximately a factor of two while preserving
the relative strength of the spectral features.  We also show for comparison the synthetic
spectrum of a hydrogen deficient model with a value of $T_{\rm eff}$ that is 1kK greater than that
of the solar abundance model.  The synthetic spectrum from the hydrogen deficient model with the
larger $T_{\rm eff}$ approximately matches both the overall flux level and the detailed structure of
the synthetic spectrum of the solar abundance model throughout the IUE wavelength range.  To a first 
approximation, a solar abundance model is interchangeable with a hydrogen deficient model that is
1 kK hotter when fitting the near UV spectrum.  For the accuracy of the fitting done here, it is 
necessary to fit models to a region of
the spectrum other than the UV to disentangle $T_{\rm eff}$ from $[\frac{A}{H}]$.

\paragraph{}

\begin{figure}
\plotone{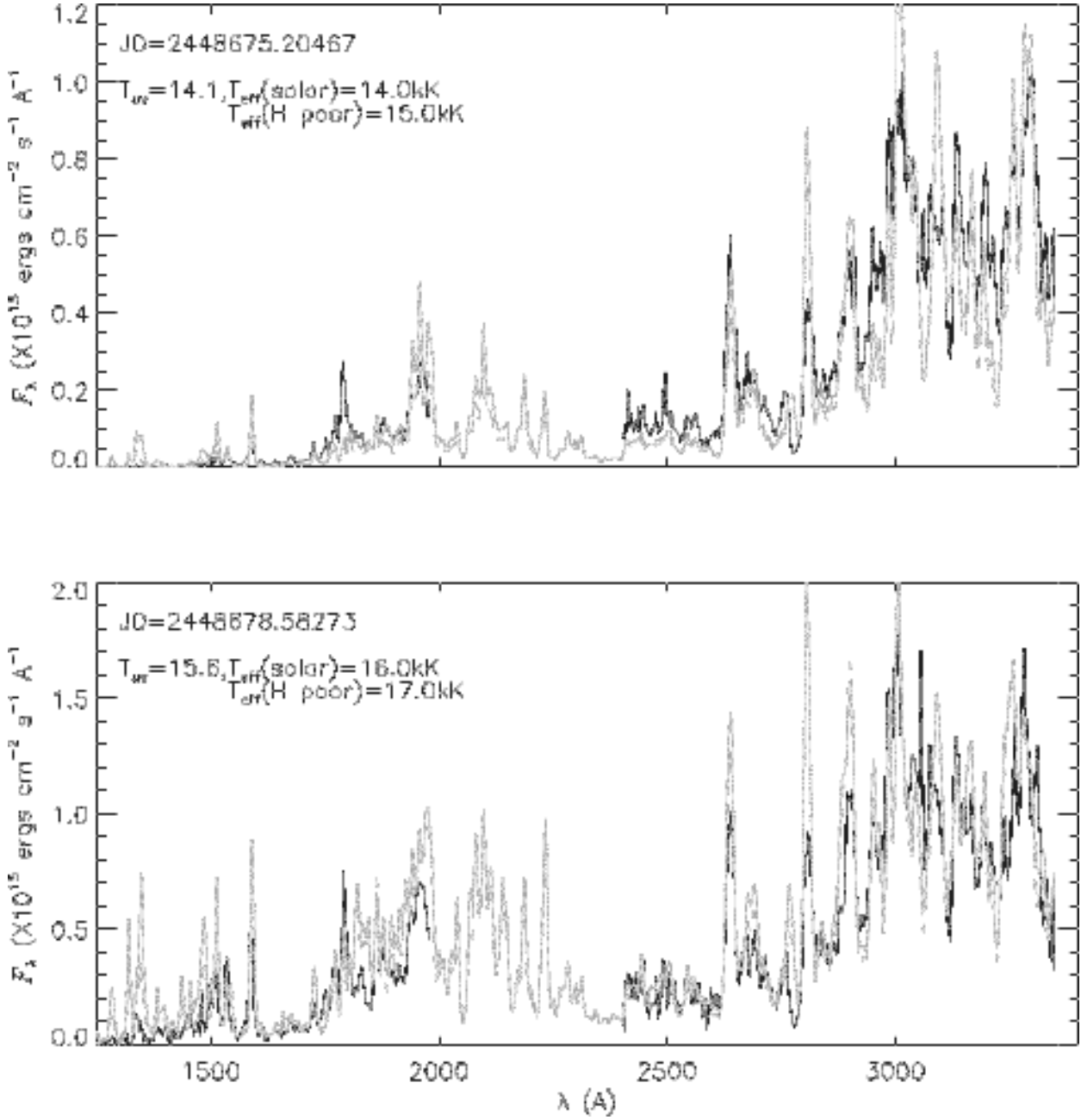}
\figcaption[f15.eps]
{Comparison of observed (darker line) and synthetic (lighter lines) spectra
at two representative times.  Lighter solid line: $[{A\over H}]=0$ (solar abundance) 
models; lighter dashed line: $[{A\over H}]=0.3$ models. 
\label{2H_plot} } 
\end{figure}

    Fig. \ref{2H_plot} shows a comparison of the observed spectrum and synthetic spectra from
the best fit solar abundance model ($T_{\rm eff}=T_{\rm UV}$) and a hydrogen deficient model
of $T_{\rm eff}=T_{\rm UV}+1$ kK at two times during the optically thick wind phase.  From a 
visual inspection, both synthetic spectra provide approximately the same goodness of fit at both times.
We conclude that the result of fitting hydrogen deficient models to the observed spectral
sequence is to shift the derived $T_{\rm eff}$ evolution upward by approximately 1 kK.  There
are no spectral features in this range that distinguish between two such models.  Therefore,
there is a degeneracy in $T_{\rm eff}$ and $n_{\rm H}$, at least within the range of parameters
explored here.  Superficially, the degeneracy arises because a larger value of $T_{\rm eff}$ 
enhances the UV flux (see Fig. \ref{surf_syn}), whereas the increased value of $[{Fe\over H}]$
suppresses the UV flux.  The net result is that increasing both in a particular proportion will lead
to an approximately similar UV flux spectrum.  A more fundamental discussion of spectral
similarity and non-uniqueness problems in atmospheric models of novae can be found in 
\citet{p&s}.        

\section{Conclusion}\label{conclusions}

We conclude that the UV radiation temperature ($T_{\rm UV}$) is a good measure of $T_{\rm eff}$ during
the early stage of the nova when the spectrum is purely that of an optically thick
wind.   We find that this stage lasts until JD2448682.8368, which is about 10 days after the
first fireball phase spectrum was taken.  
During this interval, {\tt PHOENIX} models are able to reproduce the overall shape of the low resolution
UV spectra in the IUE range.  This is significant because the nova 
photosphere increases in $T_{\rm eff}$ by $\approx 5000$ K during this
time, and the hardness of the near UV radiation field is sensitive to 
$T_{\rm eff}$ in the range from ten to twenty kK.  The models 
provide a moderately good fit to the high resolution spectra during 
the optically thick wind phase, but there are significant discrepancies
which are due to, among other things, a non-solar abundance distribution
due to nuclear processing, an inaccurate velocity law, and inhomogeneities 
in the expanding atmosphere.  

\paragraph{}

   Models in which H is depleted to a tenth of its 
solar value and in which $\frac{A}{H}$ is twice the solar value provide a similarly good fit
to the observed spectra in the UV if the value of $T_{\rm eff}$ of the model
is increased by $\approx 1$ kK.  We conclude that $[\frac{A}{H}]$ and $T_{\rm eff}$
are degenerate parameters when fitting models to the near UV spectrum.  It is necessary
to fit models to another $\lambda$ region to uniquely determine $T_{\rm eff}$ and
$[\frac{A}{H}]$.   

\paragraph{}
 
The high resolution spectrum of the fireball
phase of the nova has weaker emission features than an optically thick
wind phase spectrum of the same best fit $T_{\rm eff}$.  The fireball phase
spectrum is better fit by a supernova model, in which the atmosphere
is thinner and has a steeper $\rho(r)$ and $v(r)$ gradient.       



\acknowledgments

This work was supported in part by NSF grant
AST-9720704, NASA ATP grant NAG 5-8425 and LTSA grant NAG 5-3619, as
well as NASA/JPL grant 961582 to the University of Georgia, by NSF
 grant
AST-9731450, NASA grant NAG5-3505 and an IBM SUR grant to the
University of Oklahoma, and by NSF and NASA grants to 
Arizona State University.  PHH was supported in part by the P\^ole
Scientifique de Mod\'elisation Num\'erique at ENS-Lyon.  Some of the
calculations presented in this paper were performed on the 
the IBM SP2 and SGI Origin 2000 of the UGA UCNS, on the IBM BlueHorizon of
the San Diego Supercomputer Center (SDSC), with support from the
National Science Foundation, and on the IBM SP and Cray T3E of the NERSC with
support from the DoE.  We thank all these institutions for a generous
allocation of computer time.

\clearpage

\clearpage

\hoffset=-0.5truein

\clearpage

\clearpage

\clearpage

\clearpage

\clearpage

\clearpage

\hoffset=-0.0truein

\clearpage

\hoffset=-0.5truein

\clearpage

\hoffset=-0.5truein

\clearpage

\clearpage

\clearpage

\clearpage

\clearpage

\clearpage

\clearpage

\clearpage

\hoffset=+0.0truein
\begin{deluxetable}{lrlr}
\scriptsize
\tablecaption{Log of low resolution IUE spectra}
\tablecomments{Matching superscripts indicate approximately simultaneous LWP/SWP pairs.} 
\label{t2}
\tablecolumns{4}
\tablewidth{0pt}
\tablehead{
\multicolumn{2}{c}{LWP} & \multicolumn{2}{c}{SWP} \\
\colhead{Exp. No.} & \colhead{JD-2448000} & \colhead{Exp. No.} & \colhead{JD-2448000} }
\startdata 
 ${\rm lwp}22425^1$    &  673.40  & ${\rm swp}44030^1$    &  673.41 \\
 ${\rm lwp}22426^2$    &  673.45  & ${\rm swp}44031^2$    &  673.46 \\
 ${\rm lwp}22430^3$    &  674.40  & ${\rm swp}44039^3$    &  674.40 \\
 ${\rm lwp}22434^4$    &  675.19  & ${\rm swp}44040$      &  674.45 \\
 ${\rm lwp}22435^5$    &  675.25  & ${\rm swp}44043^4$    &  675.20 \\
 ${\rm lwp}22448$      &  676.41  & ${\rm swp}44044^5$    &  675.27 \\
 ${\rm lwp}22449^6$    &  676.45  & ${\rm swp}44050^6$    &  676.44 \\
 ${\rm lwp}22456^7$    &  677.62  & ${\rm swp}44051$      &  676.69 \\
 ${\rm lwp}22459^8$    &  678.57  & ${\rm swp}44055^7$    &  677.61 \\
 ${\rm lwp}22462^9$    &  678.86  & ${\rm swp}44056$      &  677.66 \\
 ${\rm lwp}22470$      &  680.50  & ${\rm swp}44060^8$    &  678.58 \\
 ${\rm lwp}22483^{10}$ &  682.86  & ${\rm swp}44062$      &  678.69 \\
 ${\rm lwp}22485$      &  682.94  & ${\rm swp}44064^9$    &  678.86 \\
 ${\rm lwp}22513^{11}$ &  686.38  & ${\rm swp}44073$      &  680.43 \\
 ${\rm lwp}22533^{12}$ &  688.85  & ${\rm swp}44086^{10}$ &  682.83 \\
 ${\rm lwp}22573$      &  693.43  & ${\rm swp}44102$      &  684.48 \\
 ${\rm lwp}22592^{13}$ &  695.68  & ${\rm swp}44115^{11}$ &  686.32 \\
 ${\rm lwp}22635^{14}$ &  700.49  & ${\rm swp}44130^{12}$ &  688.85 \\
 ${\rm lwp}23077$      &  752.34  & ${\rm swp}44174^{13}$ &  695.69 \\
 ${\rm lwp}23134$      &  761.76  & ${\rm swp}44193^{14}$ &  700.49 \\
 ${\rm lwp}23170$      &  767.66  & ${\rm swp}44209$      &  703.85 \\
 ${\rm lwp}23188$      &  770.31  & ${\rm swp}44233$      &  707.50 \\
 ${\rm lwp}23210$      &  772.64  & ${\rm swp}44268$      &  711.89 \\
 ${\rm lwp}23276$      &  783.24  & ${\rm swp}44305$      &  715.32 \\
 ${\rm lwp}23312$      &  790.26  & ${\rm swp}44338$      &  717.77 \\
 ${\rm lwp}23348$      &  794.42  & ${\rm swp}44377$      &  723.43 \\
 ${\rm lwp}23351$      &  795.67  & ${\rm swp}44389$      &  725.76 \\
 ${\rm lwp}23399$      &  802.58  & ${\rm swp}44439$      &  732.75 \\
 ${\rm lwp}23424$      &  807.15  & ${\rm swp}44632$      &  752.31 \\
 ${\rm lwp}23425$      &  807.20  & ${\rm swp}44634$      &  752.42 \\
 ${\rm lwp}23426$      &  807.31  & ${\rm swp}44717$      &  761.70 \\
 ${\rm lwp}23501$      &  818.17  & ${\rm swp}44761$      &  767.67 \\
 ${\rm lwp}23599$      &  833.04  & ${\rm swp}44762$      &  767.70 \\
 ${\rm lwp}23670$      &  845.01  & ${\rm swp}44790$      &  770.31 \\
 ${\rm lwp}23706$      &  851.43  & ${\rm swp}44808$      &  772.64 \\
 ${\rm lwp}23802$      &  864.30  & ${\rm swp}44901$      &  783.26 \\
 ${\rm lwp}23869$      &  873.00  & ${\rm swp}44937$      &  790.24 \\
 ${\rm lwp}24153$      &  921.60  & ${\rm swp}44970$      &  794.43 \\
\enddata
\end{deluxetable}

\clearpage

\begin{deluxetable}{lrrrrrrr}
\scriptsize
\tablecaption{Ranges of $T_{\rm eff}$ (in kK) in the model grid for which each ionization stage of each
element was treated in NLTE.}
\tablecomments{Elements in bold face have been added since the last NLTE {\tt PHOENIX} analysis
of Cygni 1992.  Those in italics have had their treatment improved since the last analysis.} 
\label{t3}
\tablecolumns{8}
\tablewidth{0pt}
\tablehead{
\colhead{Element} & \multicolumn{7}{c}{Ionization Stage} \\
\colhead{} & \colhead{\ion{}{1}} & \colhead{\ion{}{2}} & \colhead{\ion{}{3}} & \colhead{\ion{}{4}} & \colhead{\ion{}{5}} & \colhead{\ion{}{6}} & \colhead{\ion{}{7}}   } 
\startdata
{\it H}  & All &\nodata &\nodata &\nodata &\nodata &\nodata &\nodata \\
He       & All & All &\nodata &\nodata &\nodata &\nodata &\nodata \\
Li       & None & None &\nodata &\nodata &\nodata &\nodata &\nodata \\
C        & 12-24 & All & All & All &\nodata &\nodata &\nodata \\
N        & All & All & All & All & None & None  &\nodata\\
O        & All & All & All & All & None & None &\nodata\\
Ne       & All &\nodata &\nodata &\nodata &\nodata &\nodata &\nodata\\
{\it Na} & 12-24 & All & All & All & None & None  &\nodata\\
{\it Mg} & 12-24 & All & All & All & None & None  &\nodata\\
{\bf Al} & 12-24 & All & All & All & None & None  &\nodata\\
Si       & 12-24 & All & All & All & 30 & None &\nodata \\
{\bf P}  & 12-22 & 12-22 & 12-22 & 12-22 & None & None  &\nodata\\
S        & 12-24 & All & All & All & 30 & None  &\nodata\\
{\bf K}  & None  & None & None & None & None  &\nodata &\nodata\\
Ca       & 12-24 & 12-24 & All & All & All & None & None \\
Ti       & None & None  &\nodata &\nodata &\nodata &\nodata &\nodata\\
Fe       & 12-24 & All & All & All & All & None  &\nodata\\
Co       & None & None & None  &\nodata &\nodata &\nodata &\nodata\\
{\bf Ni} & None & 25-30 & 25-30 & 25-30 & 30 & None &\nodata\\
\enddata
\end{deluxetable}



\begin{thebibliography}{}
\bibitem[Castor {\it et al.}(1975)]{CAK} Castor, J.D., Abbott, D., \& Klein, R., 1975, \apj, 195, 157
\bibitem[Chochol {\it et al.}(1997)]{chochol} Chochol, D., Grygar, J., Pribullai, T., Komzik, R., 
Hric, L., \& Elkin, V. 1997, \aap, 318, 908
\bibitem[Collins(1992)]{collins} Collins, M., 1992, IAU Circ. No. 5454
\bibitem[Fitzpatrick(1999)]{fitz} Fitzpatrick, E. 1999, \pasp, 111, 63
\bibitem[Hauschildt \& Baron(1999)]{hausphx} Hauschildt, P. H. \& Baron, E., 1999, J. Comput. App. Math. 102, 41 
\bibitem[Hauschildt {\it et al.}(1994)]{ncyg2} Hauschildt, P. H., Starrfield, S., Austin, S., 
Wagner, R. M., Shore, S. N., \& Sonneborn, G., 1994, \apj, 422, 831
\bibitem[Massa and Fitzpatrick(1998)]{massa} Massa, D. \& Fitzpatrick, E. 1998, \aap, 193, 1122
\bibitem[McLaughlin (1947)]{mclaugh} McLaughlin, D., 1947, \pasp, 59, 244 
\bibitem[Payne-Gaposchkin(1957)]{payne} Payne-Gaposchkin, C., 1957, The Galactic Novae (Amsterdam:
North-Holland)
\bibitem[Pistinner and Shaviv(1996)]{p&s} Pistinner, S. \& Shaviv, G., 1996, \apj, 461, L45 
\bibitem[Pistinner {\it et al.}(1995)]{pistin} Pistinner, S., Shaviv, G., Hauschildt, P. \& Starrfield, S., 1995, \apj 451, 724
\bibitem[Shore {\it et al.}(1994)]{ncyg1} Shore, S. N., Sonneborn, G., Starrfield, S., Gonzalez-Riestra \&
R., Polidan, R. S., 1994, \apj, 421, 344 
\bibitem[Shore {\it et al.}(1997)]{nebul} Shore, S. N., Starrfield, S., Ake, T. B. \& Hauschildt, P. H.,
1997, \apj, 490, 393  
\bibitem[Short {\it et al.}(1999)]{short1} Short, C. I., Hauschildt, P. H., \& Baron, E., 1999, \apj, 525, 375
\bibitem[Starrfield {\it et al.}(1997)]{wdmass} Starrfield, S., Truran, J. W., Wiescher, M. C., \& Sparks, W. M., 1997, \mnras, 296, 502 
\end{thebibliography}
\end{document}